\documentclass[aps,prd,preprint,nofootinbib]{revtex4}
\usepackage{amsfonts}
\usepackage{mathrsfs}
\usepackage{graphicx}
\usepackage{amsmath}
\usepackage{amssymb}
\usepackage{epsfig}
\usepackage{graphicx}
\usepackage{slashed}
\usepackage{color}
\usepackage{multirow}
\usepackage{tabularx} 
\newcommand{\appropto}{\mathrel{\vcenter{
			\offinterlineskip\halign{\hfil$##$\cr
				\propto\cr\noalign{\kern2pt}\sim\cr\noalign{\kern-2pt}}}}}

\usepackage{ulem}
\usepackage{adjustbox}

\usepackage{stmaryrd}
\usepackage{amsfonts}
\usepackage{mathrsfs}
\usepackage{graphicx}
\usepackage{amsmath}
\usepackage{amssymb}
\usepackage{epsfig}
\usepackage{graphicx}
\usepackage{slashed}
\usepackage{color}
\usepackage{multirow}
\usepackage{ulem}

\usepackage[colorlinks,linkcolor=blue, anchorcolor=green,citecolor=blue]{hyperref}
\usepackage{threeparttable}
\usepackage{booktabs}
\usepackage{adjustbox}
\usepackage{float}
\usepackage{subfig}
\usepackage{braket}
\usepackage{bm}
\usepackage{BOONDOX-cal}
\usepackage[utf8]{inputenc}
\usepackage[mathscr]{eucal}
\usepackage{array}
\usepackage[capitalize]{cleveref}
\usepackage[figuresleft]{rotating}
\usepackage[justification=raggedright]{caption}

\begin{document}
\title{\boldmath Semileptonic decays of doubly charmed baryons with bag model}
\author{  Chao-Qiang Geng, Chia-Wei Liu, Aowen Zhou and Xiao Yu   \\}

\affiliation{School of Fundamental Physics and Mathematical Sciences, Hangzhou Institute for Advanced Study, UCAS, Hangzhou 310024, China\\
	University of Chinese Academy of Sciences, 100190 Beijing, China
	\vspace{0.6cm}} 


\date{\today}

\begin{abstract}
We study the semileptonic decays of $B_{cc}$ ${\rightarrow}$ $B_c\ell^+\nu_\ell$ with the bag model, where $\ell$ = $(e, \mu)$, $B_{cc}$ = $(\Xi_{cc}^{++}$, $\Xi_{cc}^+$, $\Omega_{cc}^+$),
and $ B_c$ are the singly charmed baryons with $J^P= 1/2^+$.
We obtain the  decay widths of $\Gamma(\Xi_{cc}^{++}{\rightarrow}\Xi_c^+e^+\nu_e, \Xi_c^{\prime+}e^+\nu_e, \Lambda_c^+e^+\nu_e, \Omega_c^+ e^+\nu_e) 
=(5.1\pm 0.1 , 11\pm 1, 0.34\pm 0.06, 0.76\pm 0.06)\times 10^{-14}$~GeV, 
$\Gamma(\Xi_{cc}^+\rightarrow \Xi_c^0e^+\nu_e, \Xi_c^{\prime0}e^+\nu_e , \Sigma_c^0e^+\nu_e) = (
5.1\pm 0.6, 11\pm 1, 1.5\pm 0.1) \times 10^{-14}$~GeV, 
and $\Gamma(\Omega_{cc}^+\rightarrow \Omega_c^0 e^+\nu_e, \Xi_c^0e^+\nu_e , \Xi_c^{\prime0} e^+\nu_e) = 
(22\pm 2, 0.32 \pm 0.04, 0.77\pm 0.06)\times 10^{-14}$~GeV.  We also get that
$\Gamma$($B_{cc}$ ${\rightarrow}$ $B_c\mu^+\nu_\mu$)/$\Gamma$($B_{cc}$ ${\rightarrow}$ $B_ce^+\nu_e$) = $0.97\sim 1.00$.
In addition, we discuss the $SU(3)$ flavor breaking effects, classified into three aspects: phase space differences, spectator quarks, and overlappings of the transited quarks. In particular, we show that the breaking effects are dominated by the phase space differences, which can be as large as 25\%. Explicitly, we find that $\Gamma(\Xi_{cc}^{++} \to \Lambda_c ^+ e^+ \nu _e) V_{cs}^2/\Gamma(\Xi_{cc}^{++} \to \Xi_c ^+ e^+ \nu_e )V_{cd}^2  = 1.24$, which is expected as $1$ under the exact $SU(3)$ flavor symmetry.
\end{abstract}

\maketitle

\section{Introduction}
In 2002, the SELEX collaboration reported a resonant structure in $\Lambda_c^+K^-\pi^+$ and $pD^+K^-$~\cite{SELEX:2002wqn,SELEX:2004lln}, which can be potentially caused by $\Xi_{cc}^+(3620)$. However, the same structure was not confirmed  by the FOCUS, BABAR and BELLE collaborations~\cite{Ratti:2003ez,BaBar:2006bab,Belle:2013htj}. Eventually, the long awaited evidence finally arrived in 2017 via $\Xi_{cc}^{++} \to \Lambda_c^+K^-\pi^+\pi^+$ at LHCb~\cite{LHCb:2017iph}, where the mass is 
determined to be
\begin{equation}
M_{\Xi_{cc}^{++}}=(3621.40\pm 0.72\pm 0.27\pm 0.14)~\text{MeV}.
\end{equation}
This encouraging finding was soon accompanied by the lifetime measurement of $\Xi_{cc}^{++}$~\cite{LHCb:2018zpl} as well as the observation of $\Xi_{cc}^{++}$ $\rightarrow$ $\Xi_{c}^{+} \pi^+$~\cite{LHCb:2018pcs}. One can reasonably expect much more experimental results in the future, providing  opportunities to deepen our knowledge of hadron physics.

\begin{figure}
  \centering
  \includegraphics[width=0.8\linewidth]{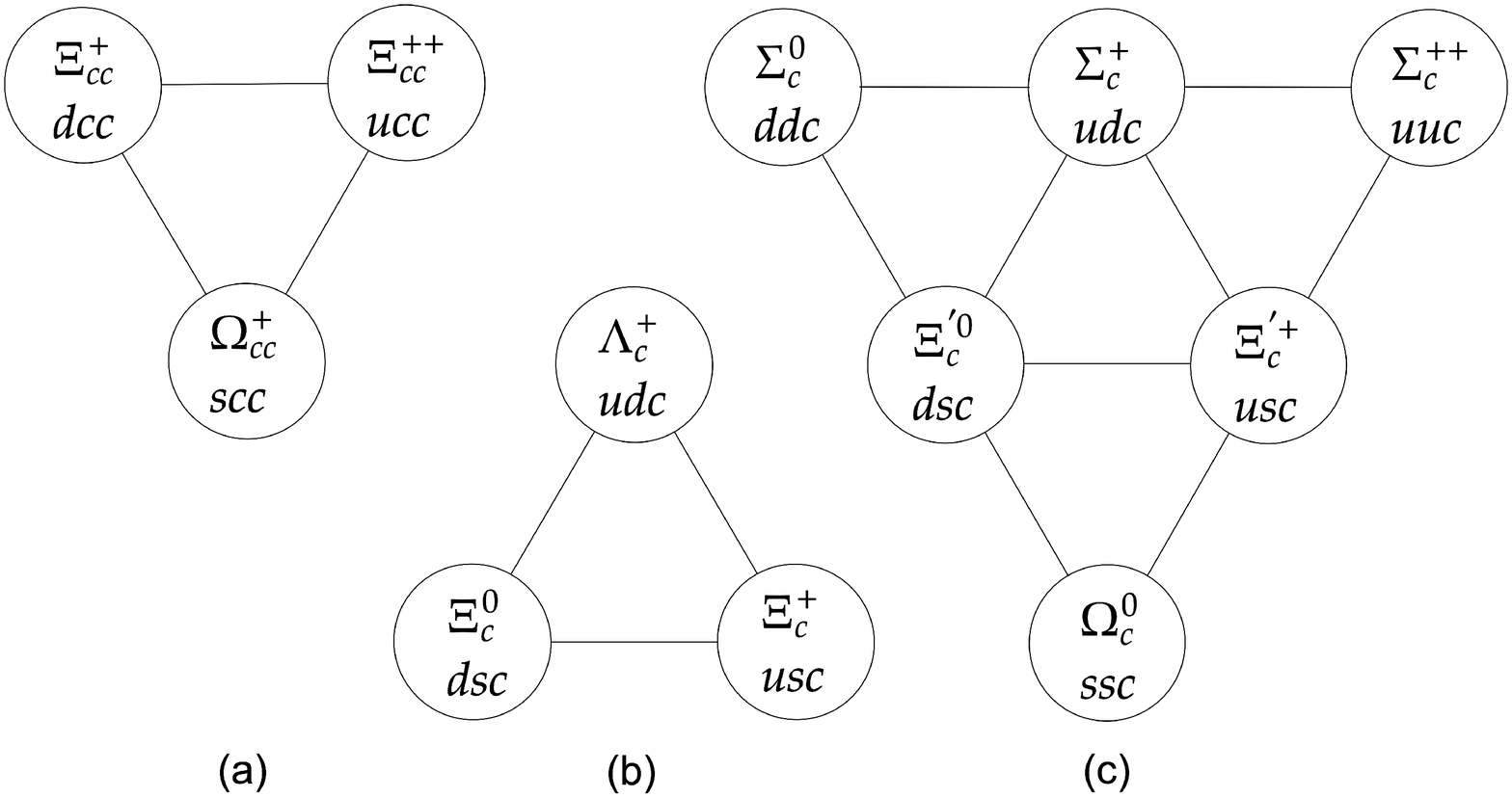}
  \caption{Quark states of the charmed baryons, where (a) represents an $SU(3)_F$ triplet with the doubly charmed baryons, while (b) and (c) correspond to the $SU(3)_F$ antitriplet and sextet with the singly charmed baryons, respectively.}
  \label{fig:su3-1}
\end{figure}

On the theoretical aspect, the low-lying charmed baryons are categorized by the representations of the flavor $SU(3)$ ($SU(3)_F$) symmetry, given in FIG.~\ref{fig:su3-1}. Under $SU(3)_F$, the doubly charmed baryons ($B_{cc}$) form a triplet, while the singly charmed baryons ($B_c$) consist of
 a antitriplet and a sextet. 
To deal with the weak decays of $B_{cc}$, lots of approaches
have been performed~\cite{Wang:2017mqp,Cerri:2018ypt,Meng:2017udf,Wang:2017azm,Gutsche:2017hux,Li:2017pxa,Guo:2017vcf,
Lu:2017meb,Xiao:2017udy,Sharma:2017txj,Ma:2017nik,Yu:2017zst,Meng:2017dni,Hu:2017dzi,Shi:2017dto,Xiao:2017dly,
Yao:2018ifh,Ozdem:2018uue,Ali:2018ifm,Dias:2018qhp,Li:2018epz,
Zhao:2018mrg,Xing:2018bqt,Ali:2018xfq,Liu:2018euh,Xing:2018lre,Dhir:2018twm,Berezhnoy:2018bde,
Jiang:2018oak,Zhang:2018llc,Meng:2018zbl,Gutsche:2018msz,Shi:2019hbf,Perez-Marcial:1989sch,Albertus:2011xz,Hu:2020mxk,Shi:2019fph,Hu:2019bqj}. 
In the diquark approach, two of the three quarks are grouped as a diquark cluster, simplifying the problem to a two-body one~\cite{Zhao:2018mrg,Xing:2018bqt,Zhu:2018epc,Ali:2018xfq,
Liu:2018euh,Xing:2018lre}.
Nonetheless, 
it is unclear 
which quarks shall form a diquark cluster. On the other hand, the problem does not exist in the MIT bag model (MBM), as a diquark cluster is unnecessary. 

The MBM describes hadrons at rest as  localized objects. Along with the bag and zero point energies, the model is suitable to explain the mass spectra. However, it becomes problematic in the decays due to the unwanted center-of-mass motion. This problem can be understood by the Heisenberg uncertainty principle, which states that a localized object can not possess a definite momentum. If we treat a bag state as a baryon at rest, the calculations will not respect the energy-momentum conservation.
The problem was tackled a few years ago by taking the linear superposition of infinite bags in Ref~\cite{Geng:2020ofy}. This approach has been applied to various decay systems~\cite{Bag1,Bag2,Bag3,Bag4,Bag5}.

This paper is organized as follows. We present the formalism of the decay branching fractions in terms of the helicity amplitudes in Sec.~\ref{FORMALISM}. In Sec.~\ref{Numerical results}, we give our numerical results and compare them with those in the literature. We summarize this work in Sec.~\ref{Conclusion}.

\section{Formalism}\label{FORMALISM}

The effective Hamiltonian for the transitions of $c\to f \ell^+ \nu$ ($f=d, s$) at the quark level is given as

\begin{equation}
\mathcal{H}_{eff} = \frac{G_F}{\sqrt{2}}V_{cf}{\bar{\ell}}
{\gamma^{\mu}}(1-\gamma_{5}) {\nu_{\ell}} \bar{f}
{\gamma}_{\mu}(1-\gamma_5)c,
\end{equation}
where $G_F$ is the Fermi constant,  and $V_{cf}$ correspond to the Cabibbo-Kobayashi-Maskawa matrix elements.
The weak transition amplitudes of the doubly charmed baryons are then given as
\begin{equation}
  \mathcal{A}(B_{cc} {\rightarrow} B_c{\ell^+}{\nu}_{\ell})=\frac{G_F}{\sqrt{2}}V_{cq}
  {\bar{\ell}}{\gamma^{\mu}}(1-\gamma_{5}) {\nu_{\ell}}{\bra{B_{c},p_f}}\bar{f}{\gamma}_{\mu}
  (1-\gamma_5)c{\ket{B_{cc},p_i}}\,,
\end{equation}
with the baryon matrix elements parameterized by 
\begin{equation}
  \begin{aligned}
  &\bra{B_{c},p_f,\lambda_f} 
  \bar{f}{\gamma}_{\mu}
  (1-\gamma_5)c{\ket{B_{cc},p_i,\lambda_i}}
  \\ 
  =&\bar{u}_{f}(p_f,\lambda_f) \left[\gamma_{\mu}f_1(q^2)-i\sigma_{\mu \nu}
  \frac{q^{\nu}}{M_i}f_2(q^2)+f_3(q^2)\frac{q_{\mu}}{M_i} \right]u_{i}(p_i,\lambda_i )
  \\
  &-\bar{u}_{f}(p_f,\lambda_f) \left[\gamma_{\mu}g_1(q^2)-i\sigma_{\mu \nu}
  \frac{q^{\nu}}{M_i}g_2(q^2)+g_3(q^2)\frac{q_{\mu}}{M_i} \right]{\gamma_5}u_{i}(p_i,\lambda_i),
  \end{aligned}
\end{equation}
where  $ f_{123}(q^2) $ and $ g_{123}(q^2) $ are the form factors, $ {\sigma^{\mu\nu}}=i
\left[{\gamma}^{\mu},{\gamma}^{\nu}\right]/2 $, $ q_\mu=p_i^\mu-p_f^\mu$, and $\lambda_{f(i)}$, $p^\mu_{f(i)}$,  $M_{f(i)}$ and $u_{f(i)}$ are the helicity, four-momentum, mass and Dirac spinor of $B_{c(c)}$, respectively.

In order to calculate the decay widths,
we introduce a set of  helicity amplitudes $ H^{V(A)}_{\lambda_f\lambda_W} $, where $\lambda_f$ 
and $\lambda_W $ represent the helicity quantum numbers of $B_c$ and the off-shell 
$ W^+ $ boson, respectively.  Relations between the helicity amplitudes and form factors
are given by~\cite{Kadeer:2005aq}
\begin{equation}
    \begin{aligned}
     & H_{\frac{1}{2}1}^V = \sqrt{2Q_-}\left(-f_1(q^2)-\frac{M_i+M_f}{M_i}f_2(q^2)\right),
     \\
     & H_{\frac{1}{2}0}^V = \frac{\sqrt{Q_-}}{q^2}\left((M_i+M_f)f_1(q^2)+\frac{q^2}{M_i}f_2(q^2)\right),
     \\
     & H_{\frac{1}{2}t}^V = \frac{\sqrt{Q_+}}{q^2}\left((M_i-M_f)f_1(q^2)+\frac{q^2}{M_i}f_3(q^2)\right),
     \\
     & H_{\frac{1}{2}1}^A = \sqrt{2Q_+}\left(g_1(q^2)-\frac{M_i-M_f}{M_i}g_2(q^2)\right),
     \\
     & H_{\frac{1}{2}0}^A = \frac{\sqrt{Q_+}}{q^2}\left(-(M_i-M_f)g_1(q^2)+\frac{q^2}{M_i}g_2(q^2)\right),
     \\
     & H_{\frac{1}{2}t}^A = \frac{\sqrt{Q_-}}{q^2}\left(-(M_i+M_f)g_1(q^2)+\frac{q^2}{M_i}g_3(q^2)\right),
    \end{aligned}
\end{equation}
where $ Q_{\pm } = (M_i\pm M_f)^2-q^2 $ and  $H_{{\lambda_f}{\lambda_W}}^{V(A)}=(-)H_{-{\lambda_f}-{\lambda_W}}^{V(A)}$.
The differential decay widths are given in
terms of the helicity amplitudes as~\cite{Geng:2019bfz,Kadeer:2005aq,Korner:1994nh}
\begin{equation}\label{decay width gamma}
  \begin{aligned}
  \partial_{q} \Gamma = & \frac{\partial \Gamma }{\partial q^2 } = 
  \frac{1}{3}\frac{G_F^2}{(2\pi)^3}|V_{fc}|^2
  \frac{(q^2-m_\ell^2)^2p}{8M_i^2q^2}\left[\left(1+
  \frac{m_\ell^2}{2q^2}\right)
  \right.
\\
  &\left.
   \left(
   \left|H_{\frac{1}{2}1}\right|^2
  +\left|H_{-\frac{1}{2}-1}\right|^2
  +\left|H_{\frac{1}{2}0}\right|^2
  +\left|H_{-\frac{1}{2}0}\right|^2
  \right)
  +\frac{3m_\ell^2}{2q^2}
  \left(
    \left|H_{\frac{1}{2}t}\right|^2
    \left|H_{-\frac{1}{2}t}\right|^2
  \right)
  \right],
  \end{aligned}
\end{equation}
where $ p = \sqrt{Q^+Q^-}/2M_{B_i} $,
$ H_{\lambda_f\lambda_W} = H_{\lambda_f\lambda_W}^V-H_{\lambda_f\lambda_W}^A $
and $m_\ell $ is the lepton mass.

In this work, we evaluate the form factors with the homogeneous bag model~(HBM)~\cite{Bag5}. The  baryon wave functions of $B_{cc}$ are given as  
\begin{equation}
|B_{cc}, \updownarrow\rangle = \int\frac{1}{2\sqrt{3} } \epsilon^{\alpha \beta \gamma} q _{a\alpha}^{\prime\dagger} (\vec{x}_1) c_{b\beta}^\dagger(\vec{x}_2) c_{c\gamma}^\dagger (\vec{x}_3) \Psi_{A_\updownarrow(ucc)}^{abc} (\vec{x}_1,\vec{x}_2,\vec{x}_3) [d^3  \vec{x}] | 0\rangle\,,
\end{equation}
where $q'=(u,d,s)$ for $B_{cc} = (\Xi_{cc}^{++}, \Xi_{cc}^+, \Omega_{cc} ^+)$, $q^{\prime\dagger} $ and  $c^\dagger$ represent the creation operators of quarks, the Latin and Greek letters stand for the Dirac spinor and color indices, and $\Psi_A$ are the spatial wave functions defined in Refs.~\cite{Bag3,Bag4, Bag5}, respectively.  On the other hand, the wave functions of 
$B_{c}$ can be found in Ref.~\cite{Bag3}. 

We choose the Breit frame to calculate the baryon matrix elements, where $B_c$ and $B_{cc}$ have the opposite velocities $\vec{v} = v\hat{z }$ and $-\vec{v}$.  
The baryon matrix elements of the current operators are then governed by 
\begin{equation}\label{2.25}
  \begin{aligned}
&\langle {B_c (\vec{v})}, \lambda_{f}| f^{\dagger}\Upsilon c(0) | B_{cc}(-\vec{v}),  \lambda_{i}\rangle
   =\mathcal{N}_{B_{c}}{\mathcal{N}_{B_{cc}}} \int d^3\vec{x}_{\triangle}\Upsilon_{fc}^{{\lambda_{f}}{\lambda_{i}}}(\vec{x}_\Delta)\prod_{q=c,q'}D_q^v
    (\vec{x}_{\triangle}),
  \\
  & D_q^v(\vec{x}_\triangle) = \frac{1}{\gamma} \int d^3\vec{x}\phi_q^\dagger\left(\vec{x}+\frac{1}{2}
    \vec{x}_\triangle\right)\phi_q \left(\vec{x}-\frac{1}{2}\vec{x}_\triangle\right)
    e^{-2iE_qvz},
  \\
  &  \Upsilon_{fc}^{{\lambda_{f}}{\lambda_{i}}}(\vec{x}_\triangle) =
      \sum_{{\lambda_q}{\lambda_c}} N^{{\lambda_{f}}{\lambda_{i}}}_{{\lambda_q}{\lambda_c}} 
      \int d^3\vec{x}
      \phi_{f{\lambda_q}}^\dagger
      \left(\vec{x}^+\right)
      S_{\vec{v}}{\Upsilon}S_{-\vec{v}}
      \phi_{c{\lambda_c}}
      \left(\vec{x}^-\right)
      e^{2i(E_{q'}+E_c)\vec{v}\cdot \vec{x}}\,,
  \end{aligned}
\end{equation}
where  $\mathcal{N}_{B_c,B_{cc}}$ are the normalization constants, $\Upsilon$ is an arbitrary
Dirac matrix,  $\vec{x}^{\pm}$ = $\vec{x}$ ${\pm}$ $\vec{x}_{\triangle}/2$,  $\phi_q$ are the bag wave functions in the MBM,  $S_{\pm v} = a_+ \pm a_- \gamma^0 \gamma^3$ with $a_\pm=\sqrt{1\pm \gamma^2}$ and $\gamma = 1/\sqrt{1-v^2}$, and $\lambda_{q,c} \in \{ \uparrow, \downarrow\}$. 
The derivations of Eq.~\eqref{2.25} and the explicit forms of $\phi_q$  are given in Ref.~\cite{Bag5}. 
The first line of Eq.~\eqref{2.25} is the total overlapping between $B_{cc}$ and $B_c$ induced by $f^\dagger\Upsilon c(0)$ at the quark level, while the second and third terms are interpreted as  :
\begin{itemize}
\item The spectator quark effects are governed by $D_{q}^v(\vec{x}_\Delta)$, which describe  the overlapping of $q$ in two bag states separated by $\vec{x}_\Delta$.
\item The quark transitions are described by $\Upsilon_{fc}^{\lambda_f\lambda_i}$, where  $N_{\lambda_q,\lambda_c} ^{\lambda_f,\lambda_i}$ are the spin-flavor overlapping coefficients.
\end{itemize}
In the heavy constituent quark limit~$(m_{u,d,c}\to \infty)$, the formalism is reduced to 
\begin{eqnarray}
&&\langle {B_c (\vec{v})}, \uparrow| f^{\dagger} c(0) | B_{cc}(-\vec{v}),  \uparrow\rangle =  \sum_{{\lambda_q}{\lambda_c}} N^{\uparrow \uparrow}_{{\lambda_q}{\lambda_c}}\,,\nonumber\\
&&\langle {B_c (\vec{v})},  \uparrow| f^{\dagger}\gamma^0\gamma^1 \gamma_5 c(0) | B_{cc}(-\vec{v}), \downarrow\rangle =  \sum_{{\lambda_q}{\lambda_c}} N^{\uparrow \downarrow}_{{\lambda_q}{\lambda_c}}\,.
\end{eqnarray}
From the angular momentum conservation, we have that
\begin{equation}
  \begin{aligned}
    &&N^{\lambda_f\lambda_i}_{\lambda_q \lambda_c} = 0~~~~~\text{for}~~\lambda_f-\lambda_i \neq \lambda_c - \lambda_q .
  \end{aligned}
\end{equation}
It states that if the  baryon spin is (un)flipped by the  operator, then the spin of the quark shall also be  (un)flipped. In addition, by the Wiger-Eckart theorem, we find that
\begin{equation}
  N^{\uparrow \uparrow}_{\lambda_q\lambda_c} =
N^{\downarrow \downarrow}_{-\lambda_q-\lambda_c}\,,~~~~N^{\uparrow \uparrow}_{\uparrow\uparrow} -   
N^{\uparrow \uparrow}_{\downarrow\downarrow}= N^{\downarrow\uparrow}_{\downarrow\uparrow} = 
N^{\uparrow\downarrow}_{\uparrow\downarrow}.
\end{equation}
Consequently, there are only two independent numbers given as 
\begin{equation}
N_{\text{unflip}} \equiv  N^{\uparrow\uparrow}_{\uparrow\uparrow} +N^{\uparrow\uparrow}_{\downarrow\downarrow}\,,~~~~N_{\text{flip}} \equiv  N^{\downarrow\uparrow}_{\downarrow\uparrow}\,,
\end{equation}
which are collected in TABLE~\ref{tab: Overlaps}.
\begin{table}[!htb]
\caption{\label{tab: Overlaps} The spin-flavor overlappings of $B_{cc}\to B_c$.}
\centering
\begin{tabular}{ccc|ccc}
  \hline\hline
$c\to s$ &~$N_{\text{unflip}}$ ~& ~$N_{\text{flip}}$~   &  ~$c\to d$~&~$N_{\text{unflip}}$~ & ~$N_{\text{flip}}$~   \\
\hline
  $\Xi^{++}_{cc}\rightarrow \Xi^+_{c}$            & $\frac{\sqrt{6}}{2}$  & $\frac{\sqrt{6}}{6}$   &  $\Xi^{++}_{cc}\rightarrow \Lambda^{+}_{c} $      & $\frac{\sqrt{6}}{2}$  & $\frac{\sqrt{6}}{6}$   \\
  $\Xi^{+}_{cc}\rightarrow \Xi^{0}_{c}$           & $\frac{\sqrt{6}}{2}$  & $\frac{\sqrt{6}}{6}$     &  $\Omega^{+}_{cc}\rightarrow \Xi^{0}_{c} $        & $\frac{\sqrt{6}}{2}$  & $\frac{\sqrt{6}}{6}$   \\
  $\Xi^{++}_{cc}\rightarrow \Xi^{\prime +}_{c}$   & $-\frac{\sqrt{2}}{2}$ & $-\frac{5 \sqrt{2}}{6}$ &  $\Xi^{++}_{cc}\rightarrow \Sigma^{+}_{c} $       & $-\frac{\sqrt{2}}{2}$ & $-\frac{5 \sqrt{2}}{6}$\\
  $\Xi^{+}_{cc}\rightarrow \Xi^{\prime 0}_{c} $    & $-\frac{\sqrt{2}}{2}$ & $-\frac{5 \sqrt{2}}{6}$&  $\Omega^{+}_{cc}\rightarrow \Xi^{\prime 0}_{c} $ & $-\frac{\sqrt{2}}{2}$ & $-\frac{5 \sqrt{2}}{6}$\\
  $\Omega^{+}_{cc}\rightarrow \Omega^{0}_{c} $     & $-1$                  & $-\frac{5}{3}$         &  $\Xi^{+}_{cc}\rightarrow \Sigma^{0}_{c} $        & $-1$                  & $-\frac{5}{3}$   \\
\hline\hline
  \end{tabular}
\end{table}

\section{Numerical results}\label{Numerical results}

The values of $V_{cf}$ are given by~\cite{Workman:2022ynf}
\begin{equation}
\begin{aligned}
 |V_{cs}|=0.987\pm 0.011,\quad |V_{cd}|=0.221\pm 0.004,
\end{aligned}
\end{equation}
while the model parameters are taken as ~\cite{Bag5}
\begin{equation}
    R=4.7\pm 0.3 \ {\rm GeV}^{-1}\,,~~~
m_c = 1.655~\text{GeV}\,,~~~m_{u,d} =0\,,~~~m_s= 0.2\pm0.1 \ {\rm GeV}\,. 
\end{equation}
Notice that $B_{cc}$ and $B_c$ have different bag radii, found to be around $4.4$ and $5.0$~GeV$^{-1}$, respectively~\cite{Hypothetic3}. However, to simplify the formalism, we take their bag radii as equal and allow them to vary from $4.4$ to $5.0$~GeV$^{-1}$.

\begin{figure}
	\centering
	\subfloat[]{\includegraphics[width=.46\textwidth]{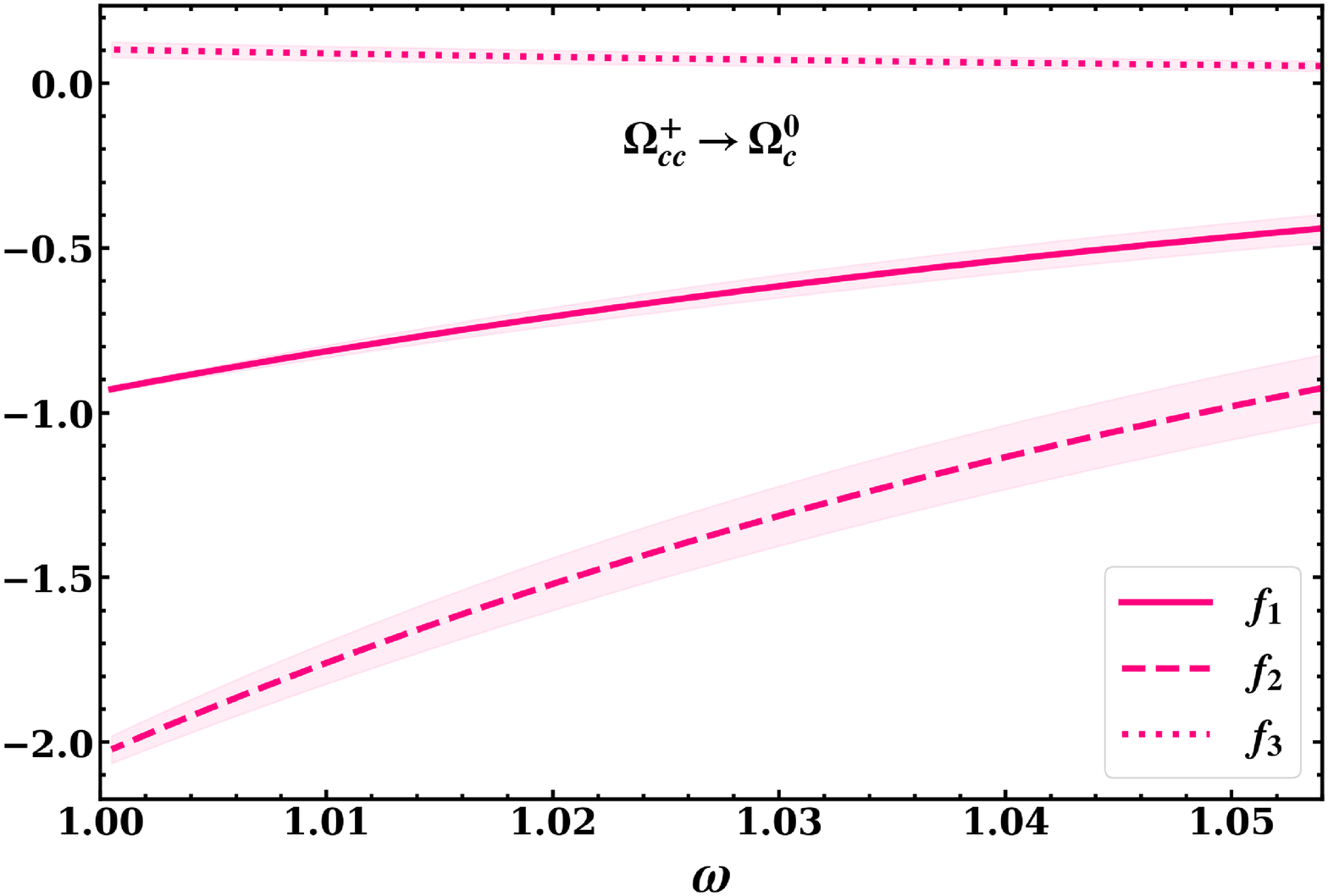}}\quad
	\subfloat[]{\includegraphics[width=.45\textwidth]{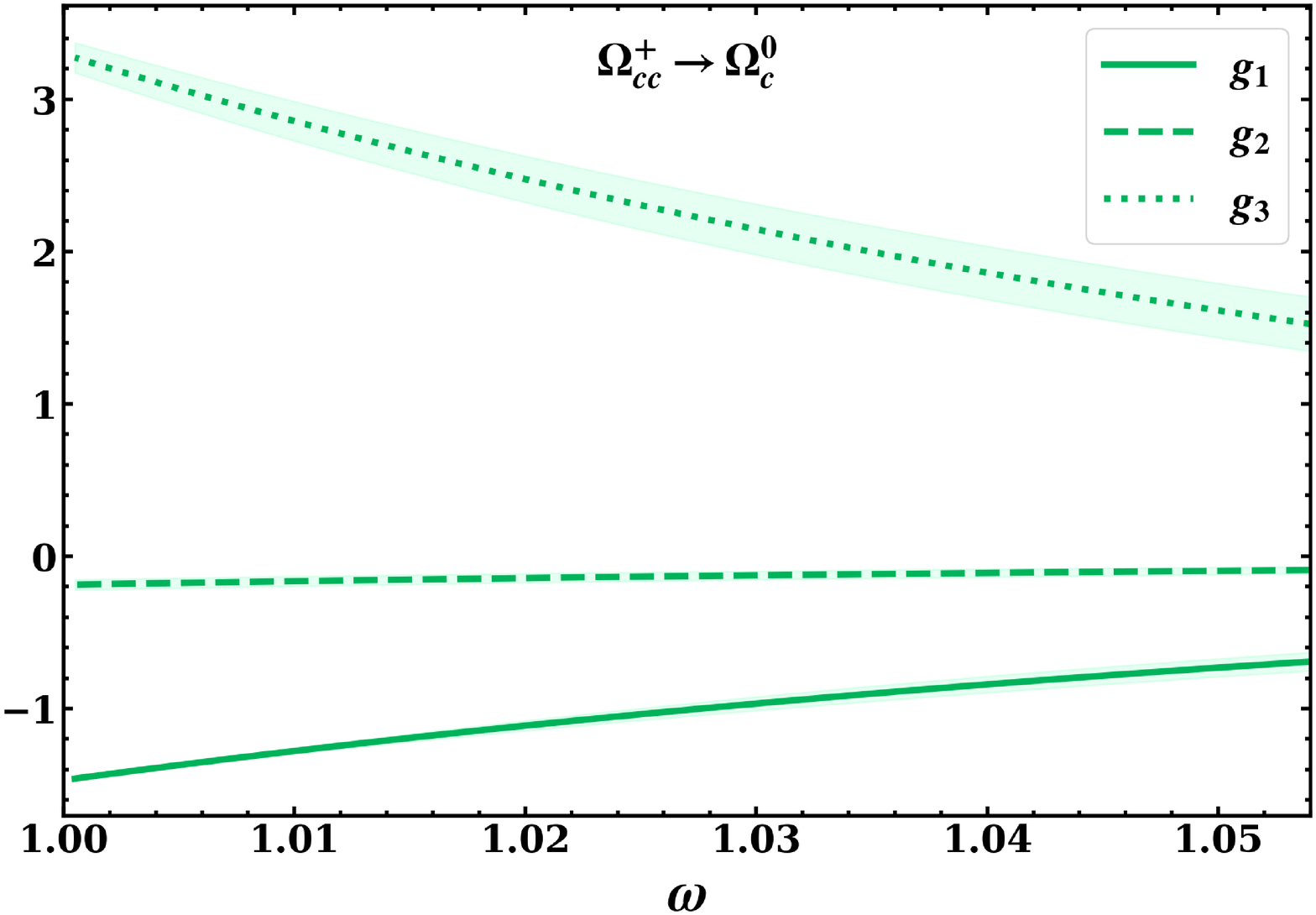}}
	\caption{The form factors of  $\Omega_{cc}^{+}\rightarrow\Omega_{c}$ as functions of $\omega$, where the center lines and bands correspond to the central values and   uncertainties, respectively.}
	\label{fig: Omega errors}
\end{figure}

To illustrate the recoil effects of the form factors, we plot those of  $\Omega_{cc}^+ \to \Omega_c$  in FIG.~\ref{fig: Omega errors}. We define
\begin{equation}
	\omega\equiv \frac{M_i^2+M_f^2-q^2}{2M_i M_f} = \frac{1+v^2}{1-v^2},
\end{equation}
so that the zero recoil point~$(q^2 =q_{max}^2 = (M_i-M_f)^2)$ corresponds to $\omega = 1$ for all the decays.  As shown in the figures, $f_3$ and $g_2$ can be taken as zero practically. The uncertainties of $f_1$ and $g_1$ are negligible at the low $q^2$ regions, and around $10\%$ at $\omega = 1.06$.  As a result, at $\omega=1$, $f_1$ and $g_1$ are not polluted by the uncertainties of the quark energies.  However, the uncertainties of $f_2$ and $g_3$  are large in all regions. 

The form factors of $\Xi_{cc}^{++}\to B_c$ at $q^2=0$, along with those in the literature, are given in Table~\ref{tab:form factors comparison}. For completeness, we also show all our calculated values of the form factors for $\Xi_{cc}^{++}\to B_{c}$ with the HBM in Table~\ref{tab:form factors}.  Note that the form factors of $\Xi_{cc}^{++} \to \Xi_c^+$  and $\Xi_{cc}^+ \to \Xi_c^0$ are the same due to the isospin symmetry. 
Compared to those in the literature, our form factors of $\Xi_{cc}^{++} \to \Lambda_c^+$ have an overall minus sign due to the convention on the baryon wave functions, which does not affect the physical quantities. Note that our results of the form factors at $q^2=0$ are significantly smaller than those in other approaches. 

\begin{table}
    \renewcommand\arraystretch{0.85}
    \caption{\label{tab:form factors comparison} 
The form factors of $\Xi_{cc}^{++}$ calculated in the HBM,  light-front quark model (LFQM), and QCD sum rule (QCDSR) at $q^2=0$.}
    \centering
\begin{tabular}{c|ccc}
    \hline
    \hline
& HBM (This work)  &LFQM~\cite{Wang:2017mqp,Shi:2019hbf} &QCDSR~\cite{Shi:2019hbf,Shi:2019fph,Hu:2019bqj}  \\
\hline
$f_{1}^{    \Xi^{++}_{cc}\rightarrow \Lambda^{+}_{c}}  $ & $0.28\pm0.05$  & $-0.79$ &$-0.59\pm0.05$\\
$f_{2}^{    \Xi^{++}_{cc}\rightarrow \Lambda^{+}_{c}}   $   & $-0.01\pm0.01$ & $0.008$ &$0.039\pm0.024$\\
$f_{3}^{    \Xi^{++}_{cc}\rightarrow \Lambda^{+}_{c}}   $   & $-0.16\pm0.02$ & - -       &$0.35\pm0.11$\\
 $g_{1}^{    \Xi^{++}_{cc}\rightarrow \Lambda^{+}_{c}}   $   & $0.09\pm0.02$  & $-0.22$&$-0.13\pm0.08$\\
$g_{2}^{    \Xi^{++}_{cc}\rightarrow \Lambda^{+}_{c}}   $   & $0.01\pm0.00$  & $0.05$&$0.037\pm0.027$\\
$g_{3}^{    \Xi^{++}_{cc}\rightarrow \Lambda^{+}_{c}}   $   & $-0.21\pm0.02$ & - -       &$0.31\pm0.09$\\
    \hline
$f_{1}^{\Xi^{++}_{cc}\rightarrow \Sigma^{+}_{c}} $   & $-0.24\pm0.01$ & $-0.47$ &$-0.35\pm0.04$\\
$f_{2}^{\Xi^{++}_{cc}\rightarrow \Sigma^{+}_{c}} $   & $-0.53\pm0.05$ & $1.04$ &$1.15\pm0.12$\\
$f^{\Xi^{++}_{cc}\rightarrow \Sigma^{+}_{c}}_{3} $   & $0.03\pm0.00$  & - -       &$-1.40\pm0.39$\\ 
$g^{\Xi^{++}_{cc}\rightarrow \Sigma^{+}_{c}}_{1} $   & $-0.37\pm0.05$ & $-0.62$& $-0.23\pm0.06$\\
$g_{2}^{\Xi^{++}_{cc}\rightarrow \Sigma^{+}_{c}} $   & $-0.05\pm0.00$ & $0.05$&$-0.26\pm0.15$\\
$g_{3}^{\Xi^{++}_{cc}\rightarrow \Sigma^{+}_{c}} $   & $0.89\pm0.06$  & - -         &$2.68\pm0.39$\\
    \hline 
    \hline
    \end{tabular}
\end{table}

\begin{table}[!htb] 
  \caption{\label{tab:form factors}  The form factors from the HBM with  $\Xi_c^{(\prime)} =(\Xi_c^{(\prime)+},\Xi_c^{(\prime) 0})$ for $\Xi_{cc} = (\Xi_{cc}^{++}, \Xi_{cc} ^+)$. }
  \centering
  \begin{threeparttable} 
  \resizebox{\textwidth}{!}
  {\begin{tabular}{c|p{4cm}<{\centering}|p{4cm}<{\centering}|c|p{4cm}<{\centering}|p{4cm}<{\centering}}
  \hline\hline 
  &$q^2=0$  &$ q^2 = q_{max}^2$   & & $q^2=0$  & $q^2 = q_{max}^2$  \\
  \hline
  $f_{1}^{\Xi_{cc}\rightarrow \Xi _{c}}$   & $0.45\pm0.05$  & $1.36\pm0.00$  & $g_{1}^{\Xi_{cc}\rightarrow \Xi _{c}}$  & $0.14\pm0.02$  & $0.38\pm0.00$ \\
  $f_{2}^{\Xi _{cc}\rightarrow \Xi _{c}}$   & $-0.03\pm0.00$ & $-0.15\pm0.01$ & $g_{2}^{\Xi _{cc}\rightarrow \Xi _{c}}$  & $0.01\pm0.00$  & $0.03\pm0.01$ \\
  $f_{3}^{\Xi _{cc}\rightarrow \Xi _{c}}$   & $-0.20\pm0.04$ & $-0.56\pm0.03$ & $g_{3}^{\Xi _{cc}\rightarrow \Xi _{c}}$  & $-0.28\pm0.04$ & $-0.79\pm0.02$\\
  \hline
  $f_{1}^{\Xi _{cc}\rightarrow \Xi^{\prime }_{c}}$   & $-0.31\pm0.03$ & $-0.70\pm0.00$ & $g_{1}^{\Xi _{cc}\rightarrow \Xi^{\prime }_{c}}$  & $-0.49\pm0.04$ & $-1.10\pm0.01$ \\
  $f_{2}^{\Xi _{cc}\rightarrow \Xi^{\prime +}_{c}}$   & $-0.63\pm0.07$ & $-1.48\pm0.02$ & $g_{2}^{\Xi _{cc}\rightarrow \Xi^{\prime }_{c}}$  & $-0.06\pm0.02$ & $-0.13\pm0.02$ \\
  $f_{3}^{\Xi _{cc}\rightarrow \Xi^{\prime +}_{c}}$   & $0.03\pm0.01$  & $0.06\pm0.02$  & $g_{3}^{\Xi _{cc}\rightarrow \Xi^{\prime }_{c}}$  & $1.00\pm0.12$  & $2.31\pm0.05$ \\
  \hline
  $f_{1}^{\Omega^{+}_{cc}\rightarrow \Omega^{0}_{c}}$    & $-0.48\pm0.04$  & $-0.99\pm0.00$ & $g_{1}^{\Omega^{+}_{cc}\rightarrow \Omega^{0}_{c}}$    & $-0.69\pm0.06$ & $-1.56\pm0.02$ \\
  $f_{2}^{\Omega^{+}_{cc}\rightarrow \Omega^{0}_{c}}$    & $-0.93\pm0.10$  & $-2.17\pm0.03$ & $g_{2}^{\Omega^{+}_{cc}\rightarrow \Omega^{0}_{c}}$    & $-0.09\pm0.02$ & $-0.20\pm0.04$ \\
  $f_{3}^{\Omega^{+}_{cc}\rightarrow \Omega^{0}_{c}}$    & $0.05\pm0.02$   & $0.11\pm0.02$  & $g_{3}^{\Omega^{+}_{cc}\rightarrow \Omega^{0}_{c}}$    & $1.52\pm0.18$  & $3.51\pm0.08$ \\
  \hline
  \hline
  $f_{1}^{\Xi^{++}_{cc}\rightarrow \Lambda^{+}_{c}}$   & $0.28\pm0.05$  & $1.42\pm0.02$  & $g_{1}^{\Xi^{++}_{cc}\rightarrow \Lambda^{+}_{c}}$  & $0.09\pm0.02$  & $0.37\pm0.00$ \\
  $f_{2}^{\Xi^{++}_{cc}\rightarrow \Lambda^{+}_{c}}$   & $-0.01\pm0.01$  & $-0.16\pm0.01$ & $g_{2}^{\Xi^{++}_{cc}\rightarrow \Lambda^{+}_{c}}$  & $0.01\pm0.00$  & $0.02\pm0.01$ \\
  $f_{3}^{\Xi^{++}_{cc}\rightarrow \Lambda^{+}_{c}}$   & $-0.16\pm0.02$  & $-0.73\pm0.06$  & $g_{3}^{\Xi^{++}_{cc}\rightarrow \Lambda^{+}_{c}}$ & $-0.21\pm0.02$  & $-0.91\pm0.05$ \\
  \hline
  $f_{1}^{\Omega^{+}_{cc}\rightarrow \Xi^{0}_{c}}$   & $0.33\pm0.05$  & $1.41\pm0.02$  & $g_{1}^{\Omega^{+}_{cc}\rightarrow \Xi^{0}_{c}}$   & $0.10\pm0.02$   & $0.37\pm0.00$ \\
  $f_{2}^{\Omega^{+}_{cc}\rightarrow \Xi^{0}_{c}}$   & $-0.01\pm0.01$  & $-0.11\pm0.02$ & $g_{2}^{\Omega^{+}_{cc}\rightarrow \Xi^{0}_{c}}$  & $0.01\pm0.00$   & $0.04\pm0.01$ \\
  $f_{3}^{\Omega^{+}_{cc}\rightarrow \Xi^{0}_{c}}$   & $-0.20\pm0.02$  & $-0.75\pm0.06$  & $g_{3}^{\Omega^{+}_{cc}\rightarrow \Xi^{0}_{c}}$ & $-0.26\pm0.03$  & $-0.99\pm0.06$ \\
  \hline
  $f_{1}^{\Xi^{++}_{cc}\rightarrow \Sigma^{+}_{c}}$   & $-0.24\pm0.01$ & $-0.70\pm0.00$  & $g_{1}^{\Xi^{++}_{cc}\rightarrow \Sigma^{+}_{c}}$ & $-0.37\pm0.05$ & $-1.08\pm0.01$ \\
  $f_{2}^{\Xi^{++}_{cc}\rightarrow \Sigma^{+}_{c}}$   & $-0.53\pm0.05$ & $-1.65\pm0.7$ & $g_{2}^{\Xi^{++}_{cc}\rightarrow \Sigma^{+}_{c}}$  & $-0.05\pm0.00$ & $-0.16\pm0.03$ \\
  $f_{3}^{\Xi^{++}_{cc}\rightarrow \Sigma^{+}_{c}}$   & $0.03\pm0.00$ & $0.09\pm0.02$  & $g_{3}^{\Xi^{++}_{cc}\rightarrow \Sigma^{+}_{c}}$  & $0.89\pm0.06$ & $2.71\pm0.16$ \\
  \hline
  $f_{1}^{\Omega^{+}_{cc}\rightarrow \Xi^{\prime 0}_{c}}$   & $-0.24\pm0.03$ & $-0.70\pm0.01$  & $g_{1}^{\Omega^{+}_{cc}\rightarrow \Xi^{\prime 0}_{c}}$ & $-0.37\pm0.05$ & $-1.08\pm0.01$ \\
  $f_{2}^{\Omega^{+}_{cc}\rightarrow \Xi^{\prime 0}_{c}}$   & $-0.56\pm0.05$ & $-1.71\pm0.07$ & $g_{2}^{\Omega^{+}_{cc}\rightarrow \Xi^{\prime 0}_{c}}$  & $-0.06\pm0.00$ & $-0.17\pm0.03$ \\
  $f_{3}^{\Omega^{+}_{cc}\rightarrow \Xi^{\prime 0}_{c}}$   & $0.04\pm0.00$ & $0.11\pm0.02$  & $g_{3}^{\Omega^{+}_{cc}\rightarrow \Xi^{\prime 0}_{c}}$  & $0.97\pm0.07$ & $2.92\pm0.18$ \\
  \hline
  $f_{1}^{\Xi^{+}_{cc}\rightarrow \Sigma^{0}_{c}}$   & $-0.38\pm0.05$  & $-0.95\pm0.03$  & $g_{1}^{\Xi^{+}_{cc}\rightarrow \Sigma^{0}_{c}}$  & $-0.52\pm0.07$ & $-1.47\pm0.05$ \\
  $f_{2}^{\Xi^{+}_{cc}\rightarrow \Sigma^{0}_{c}}$   & $-0.75\pm0.07$  & $-2.23\pm0.00$ & $g_{2}^{\Xi^{+}_{cc}\rightarrow \Sigma^{0}_{c}}$   & $-0.07\pm0.00$ & $-0.21\pm0.03$ \\
  $f_{3}^{\Xi^{+}_{cc}\rightarrow \Sigma^{0}_{c}}$   & $0.05\pm0.01$  & $0.12\pm0.03$  & $g_{3}^{\Xi^{+}_{cc}\rightarrow \Sigma^{0}_{c}}$   & $1.26\pm0.09$  & $3.67\pm0.07$ \\  
  \hline
  \hline
  \end{tabular}}
  \end{threeparttable}
\end{table}

\begin{figure}[htbp]
	\centering  
    \subfloat[]{\includegraphics[width=0.45\textwidth]{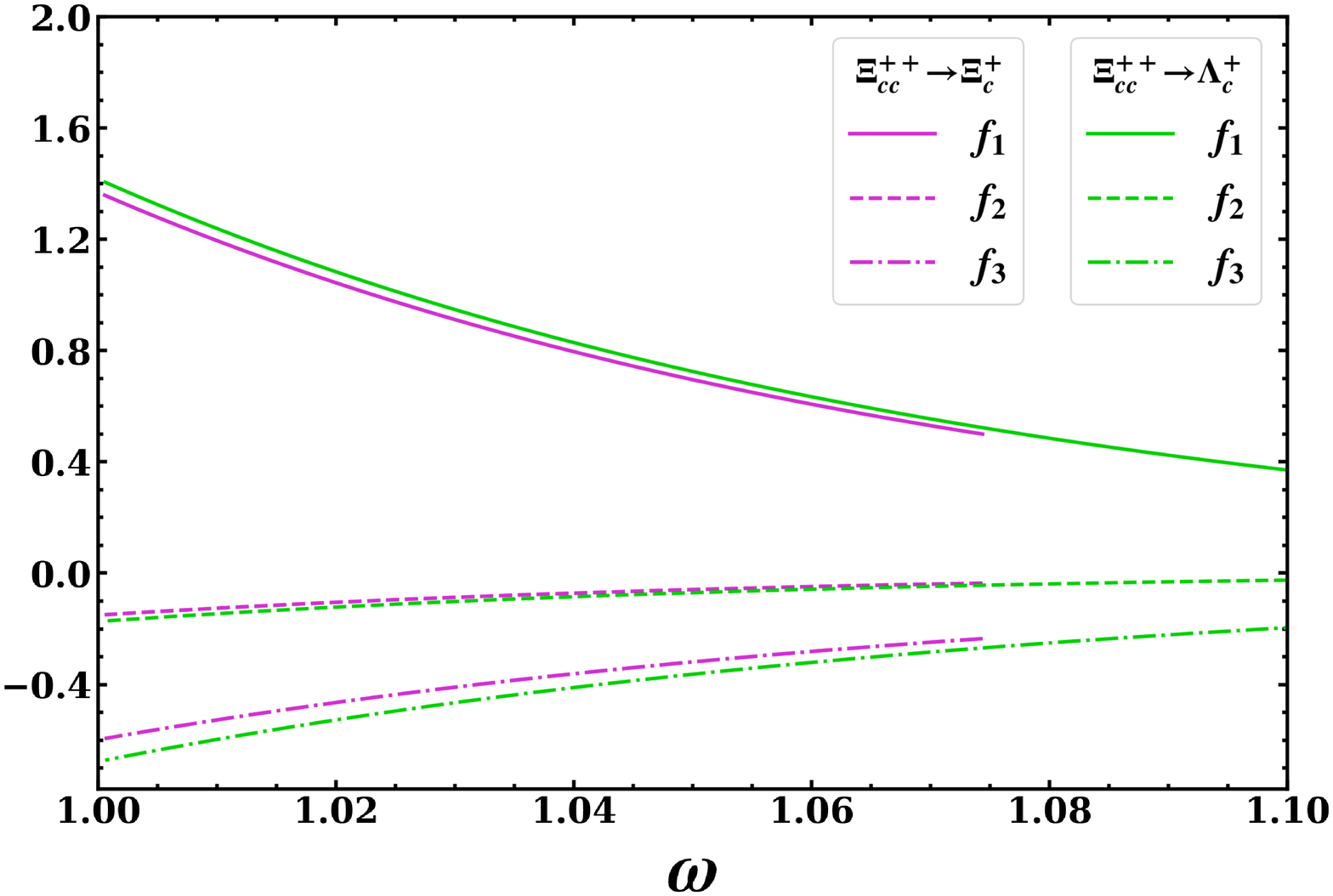}\label{fig:1-a}}\quad
	\subfloat[]{\includegraphics[width=0.45\textwidth]{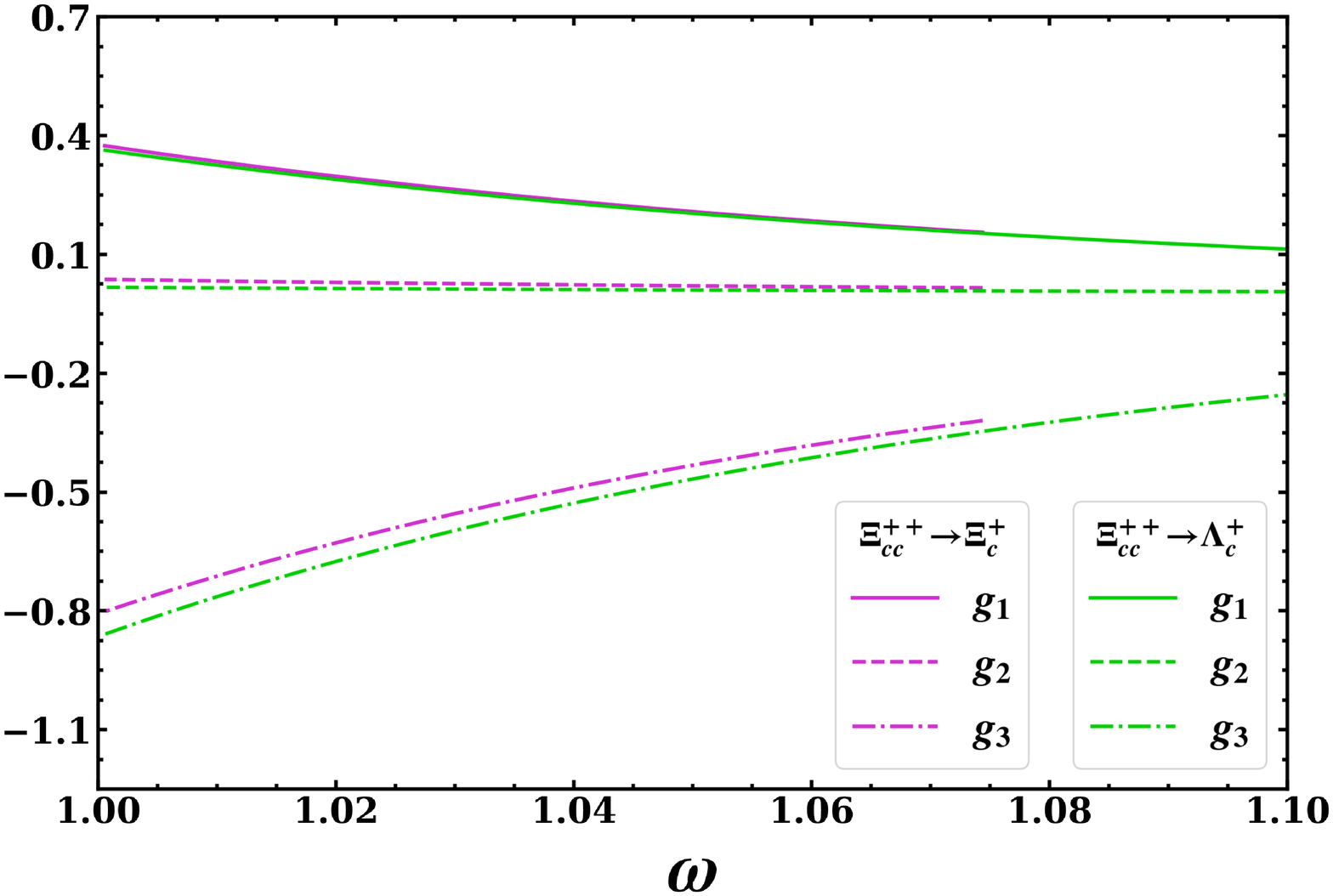}\label{fig:1-b}}
	\\
	\subfloat[]{\includegraphics[width=0.45\linewidth]{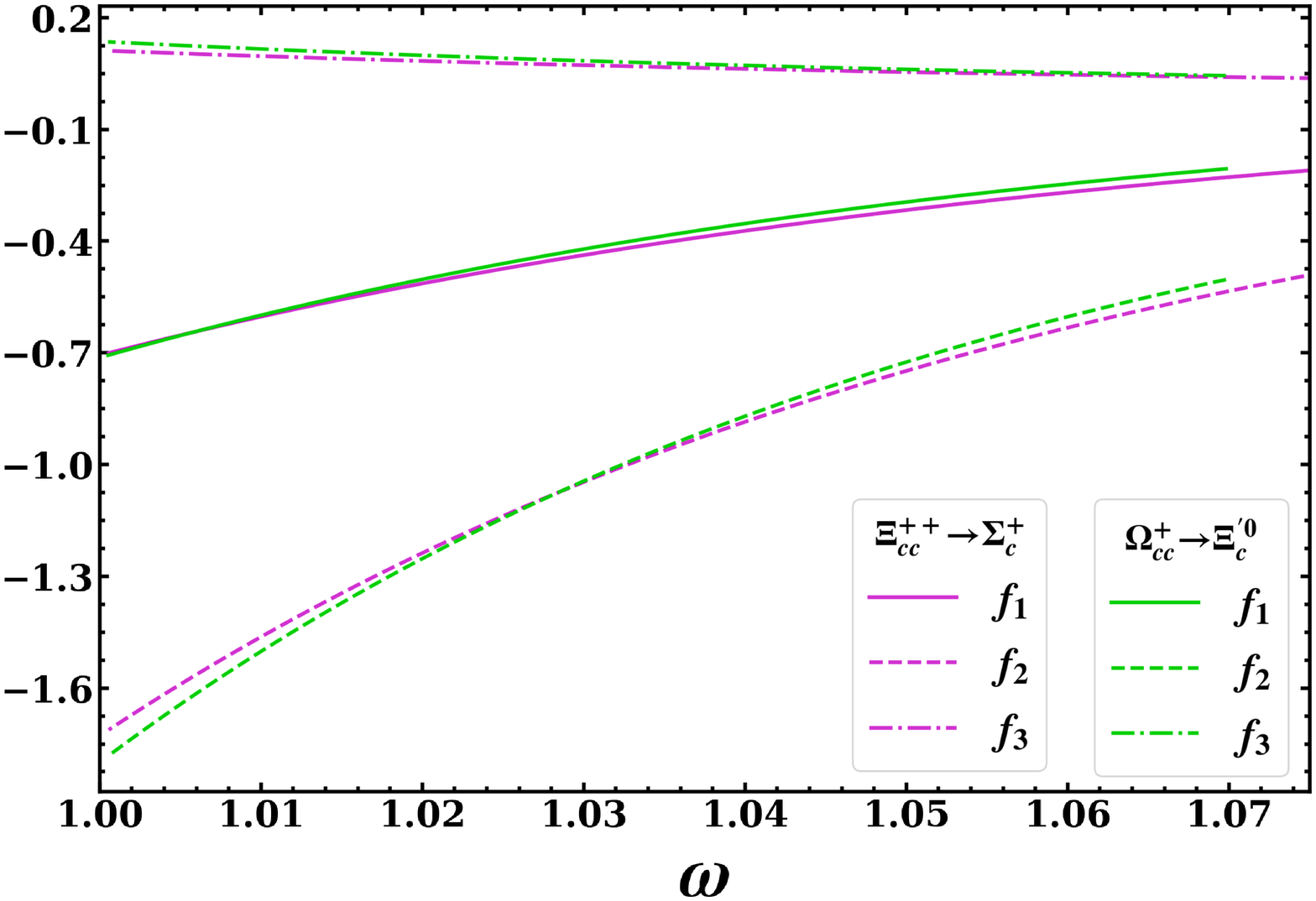}\label{fig:1-c}}\quad
	\subfloat[]{\includegraphics[width=0.45\linewidth]{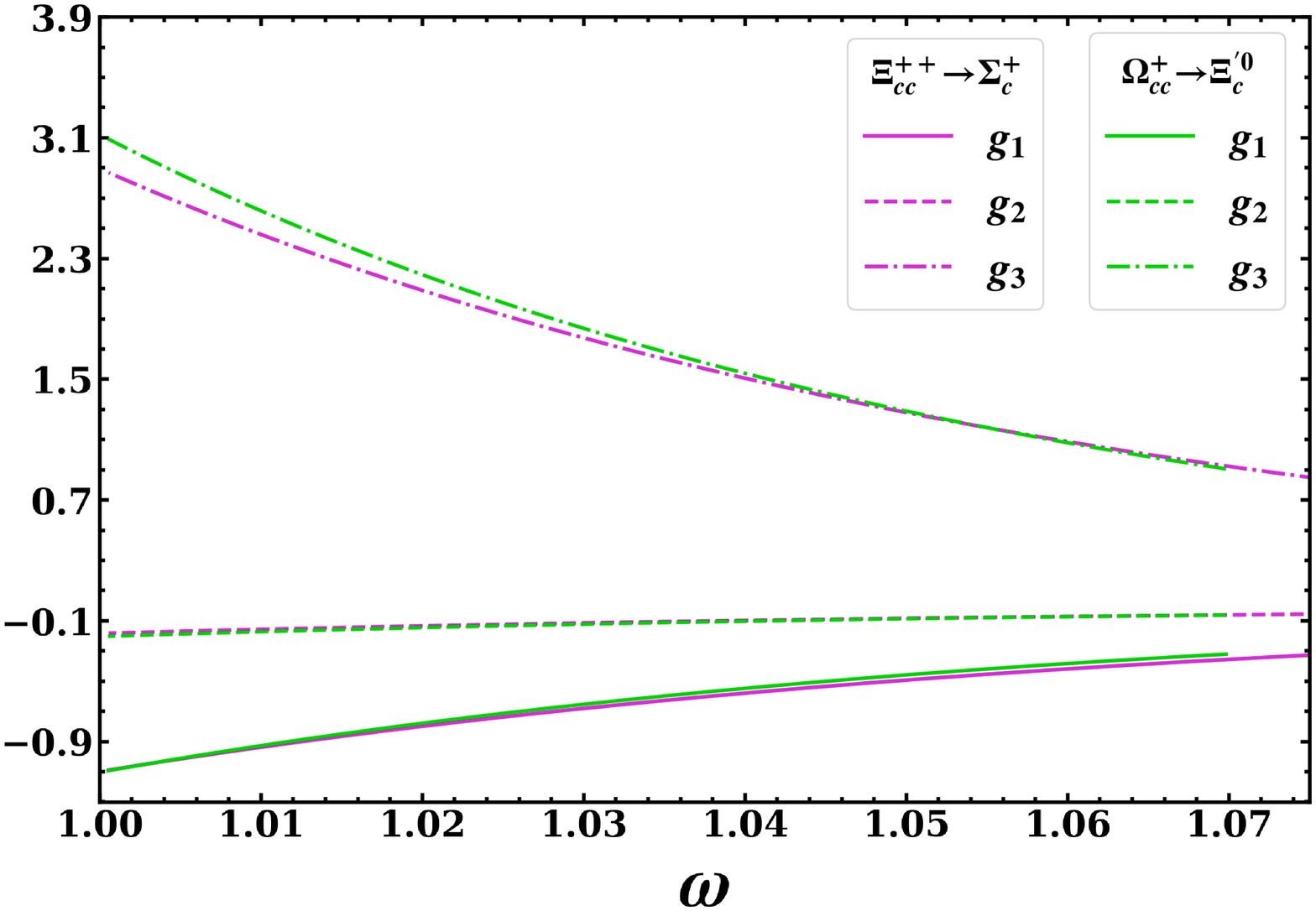}\label{fig:1-d}}
	\caption{\label{fig:1} The $\omega$ dependencies of the	$B_{cc}\rightarrow B_{c}$ form factors.}
\end{figure}

In FIGs.~\ref{fig:1-a} and \ref{fig:1-b}, we compare the form factors of 
$\Xi_{cc}^{++} \to \Xi_c^+$ and $  \Xi_{cc}^{++} \to \Lambda_c^+$, corresponding to 
$c\to s $ and $c\to d$ transitions, respectively. These form factors shall be identical in the limit of the $SU(3)_F$ symmetry. The figures show that for a fixed value of $\omega$, the form factors are approximately the same. Explicitly, their values deviate by less than $13 \%$. However, the phase space of the $c \to s $ transition is about $30\%$  smaller than the one of $c \to d$. As shown in the figures, the form factors of $\Xi_{cc}^{++} \to \Xi_c^+$ with $\omega > 1.07$ are missing as they correspond to the region of $q^2<0$. 

On the other hand, to examine the spectator effects, we plot  the results  of $\Xi_{cc}^{++} \to \Sigma_c^+$ and $\Omega_{cc}^+ \to \Xi_c^0$  in FIGs.~\ref{fig:1-c} and \ref{fig:1-d}, corresponding to the $c\to d $ transition
with $(c,u)$ and $(c,s)$ as  the spectator quarks, respectively. We find that the form factors  deviate less than $11\%$ between the two types of transition. We conclude that the form factors well respect the $SU(3)_F$ symmetry if one uses the variables of $\omega$ instead of $q^2$. 

The total decay widths of $\Gamma$ are computed by integrating Eq.~\eqref{decay width gamma}.
To further examine the results, we decompose the decay widths into four fragments, given by
\begin{equation}
    \begin{aligned}
        P_1&=\frac{1}{\Gamma}\int_{m_e^2}^{\frac{1}{4}q_{max}^2}        \partial_q \Gamma d q^2\,,~~~P_2&=\frac{1}{\Gamma}\int_{\frac{1}{4}q_{max}^2}^{\frac{1}{2}q_{max}^2}        \partial_q \Gamma d q^2,\\
        P_3&=\frac{1}{\Gamma}\int_{\frac{1}{2}q_{max}^2}^{\frac{3}{4}q_{max}^2}        \partial_q \Gamma d q^2\,,~~~ P_4&=\frac{1}{\Gamma}\int_{\frac{3}{4}q_{max}^2}^{q_{max}^2}        \partial_q \Gamma d q^2,\\
    \end{aligned}
\end{equation}
with their values listed in Table~\ref{tab: Decay width in several parts}, respectively. The uncertainties of $P_i$ are tiny compared to the total branching fractions  due to the correlations. In addition, we find that except for $\Xi_{cc}^{++} \to \Xi_c^+ e^+ \nu_e$, the values of $P_1$ are the smallest among the fragments. In contrast to the others,  the decay width of 
$\Xi_{cc}^{++} \to \Xi_c^+ e^+ \nu_e$ distributes  smoothly among the four regions.

\begin{table}
  \caption{ The total and fragmentary decay widths.}
  \label{tab: Decay width in several parts}
  \centering
  
  \begin{threeparttable} 
  \resizebox{\textwidth}{!}
  {\begin{tabular}{lp{3.5cm}<{\centering}p{3cm}<{\centering}p{3cm}<{\centering}p{3cm}<{\centering}p{3cm}<{\centering}}
\hline
\hline
channels&$\Gamma \times 10^{14} ~\text{GeV}^{-1}$ &$P_1$&$P_2$&$P_3$&$P_4$\\
  \hline 
  $\Xi^{++}_{cc}\rightarrow \Xi^+_{c}e^+ \nu_e$           & $5.11\pm 0.64$ & $0.25\pm 0.02$ & $0.29\pm 0.01$ & $0.29\pm 0.01$ & $0.17\pm 0.02$\\
  
  $\Xi^{++}_{cc}\rightarrow \Xi^{\prime +}_{c}e^+ \nu_e$  & $10.9\pm 0.8$& $0.14\pm 0.01$ & $0.24\pm 0.01$ & $0.33\pm 0.00$ & $0.29\pm 0.02$\\
  
  $\Omega^{+}_{cc}\rightarrow \Omega^{0}_{c}e^+ \nu_e$    & $22.1\pm 1.6$& $0.14\pm 0.01$ & $0.24\pm 0.01$ & $0.33\pm 0.00$ & $0.29\pm 0.02$\\
  \cline{1-6}
  $\Xi^{++}_{cc}\rightarrow \Lambda^{+}_{c}e^+ \nu_e$     & $0.34\pm 0.06$ & $0.17\pm 0.03$ & $0.26\pm 0.01$ & $0.32\pm 0.00$ & $0.24\pm 0.04$\\
  
  $\Xi^{++}_{cc}\rightarrow \Sigma^{+}_{c}e^+ \nu_e$      & $0.76\pm 0.06$ & $0.12\pm 0.02$ & $0.21\pm 0.01$ & $0.34\pm 0.00$ & $0.34\pm 0.03$\\
  
  $\Xi^{+}_{cc}\rightarrow \Sigma^{0}_{c}e^+ \nu_e$       & $1.52\pm 0.12$ & $0.12\pm 0.02$ & $0.21\pm 0.02$ & $0.34\pm 0.00$ & $0.34\pm 0.03$\\
  \hline
  \hline
  \bottomrule
  \end{tabular}}
  \end{threeparttable}
\end{table}

\begin{figure}[htb!]
    \centering
    \subfloat[]{\includegraphics[width=.45\textwidth]{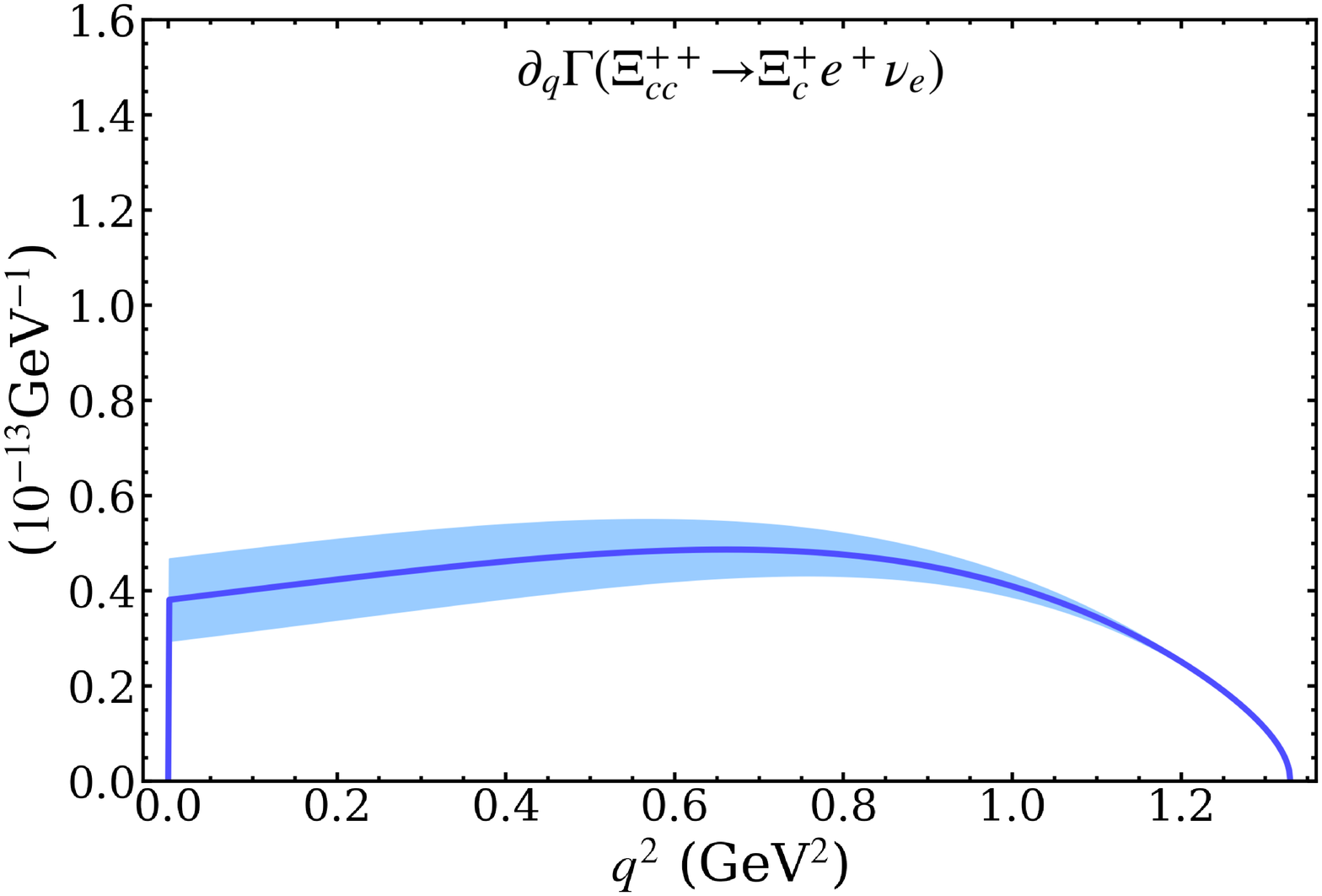}\label{fig: Omega decay widths-1}}\quad
    \subfloat[]{\includegraphics[width=.45\textwidth]{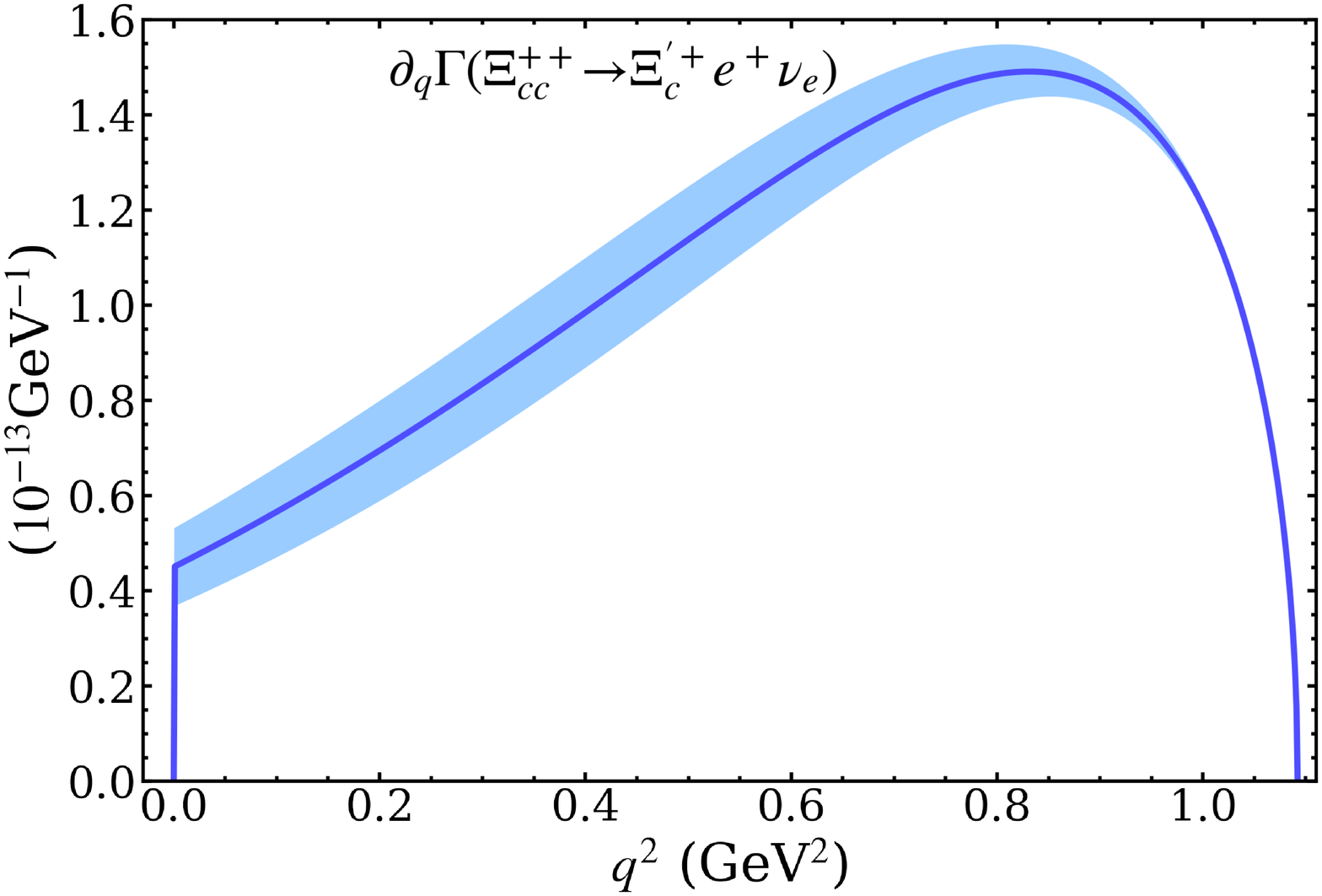}\label{fig: Omega decay widths-2}}
    \\
    \subfloat[]{\includegraphics[width=.45\textwidth]{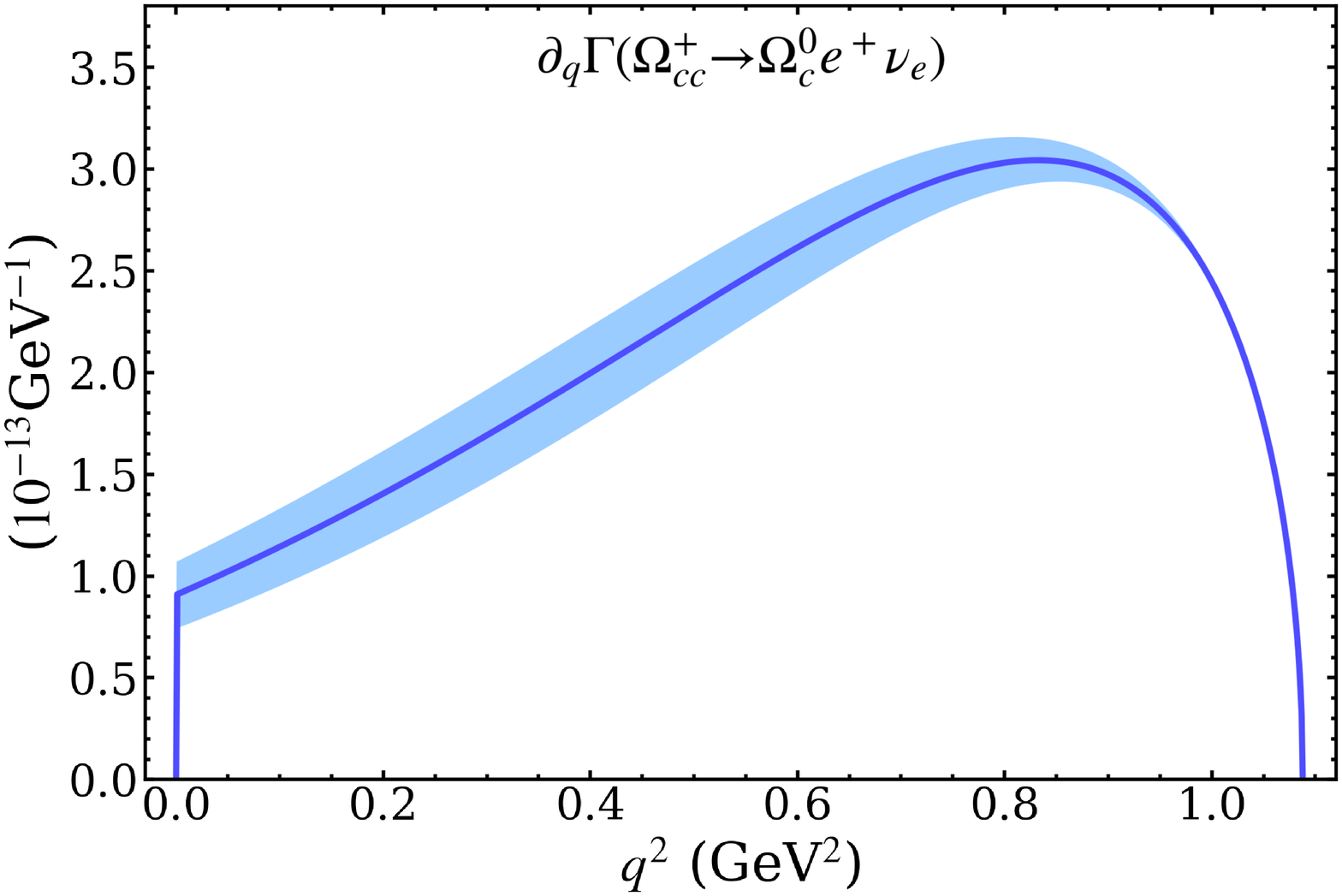}\label{fig: Omega decay widths-3}}\quad
    \subfloat[]{\includegraphics[width=.45\textwidth]{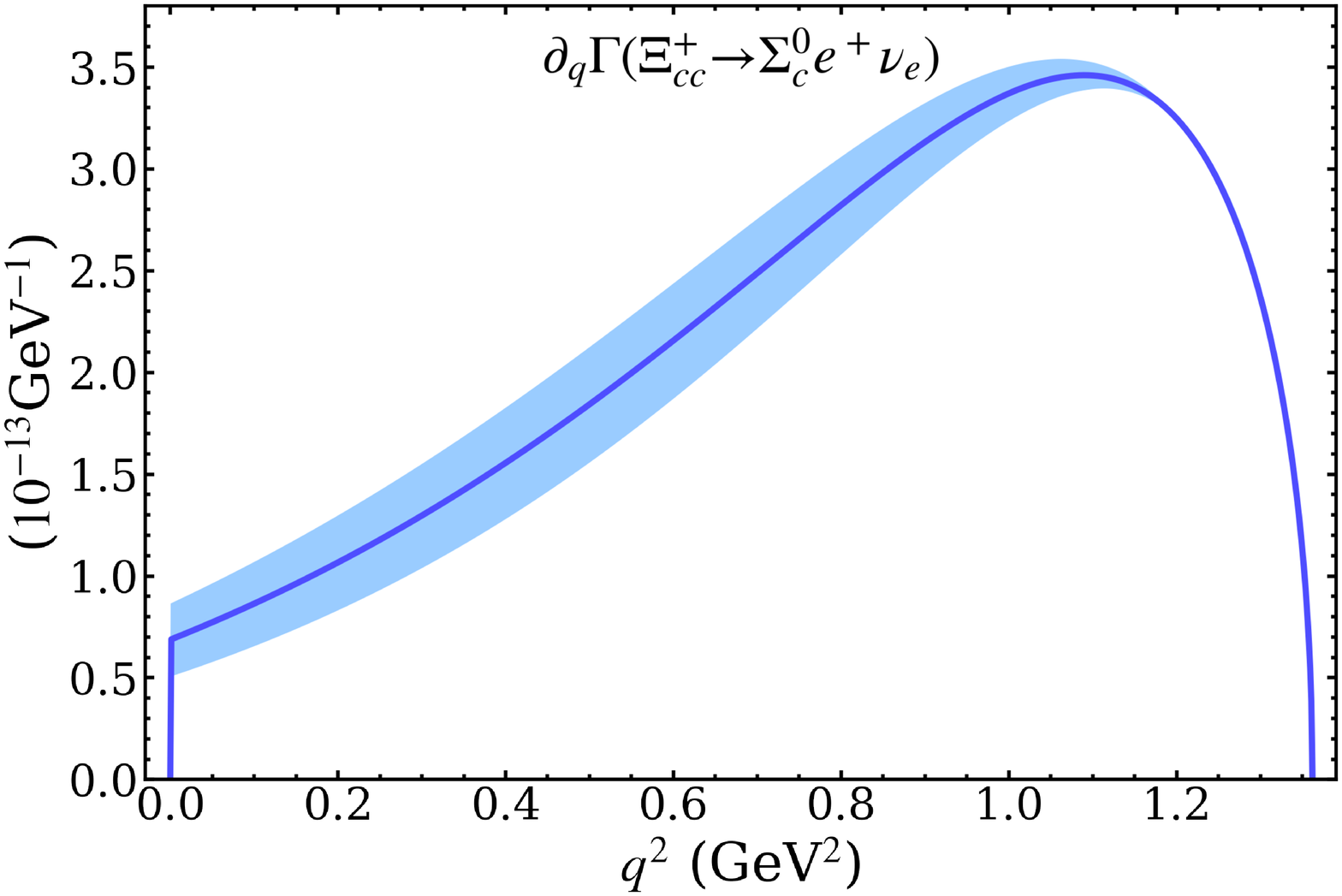}\label{fig: Omega decay widths-4}}
    \\
    \subfloat[]{\includegraphics[width=.45\textwidth]{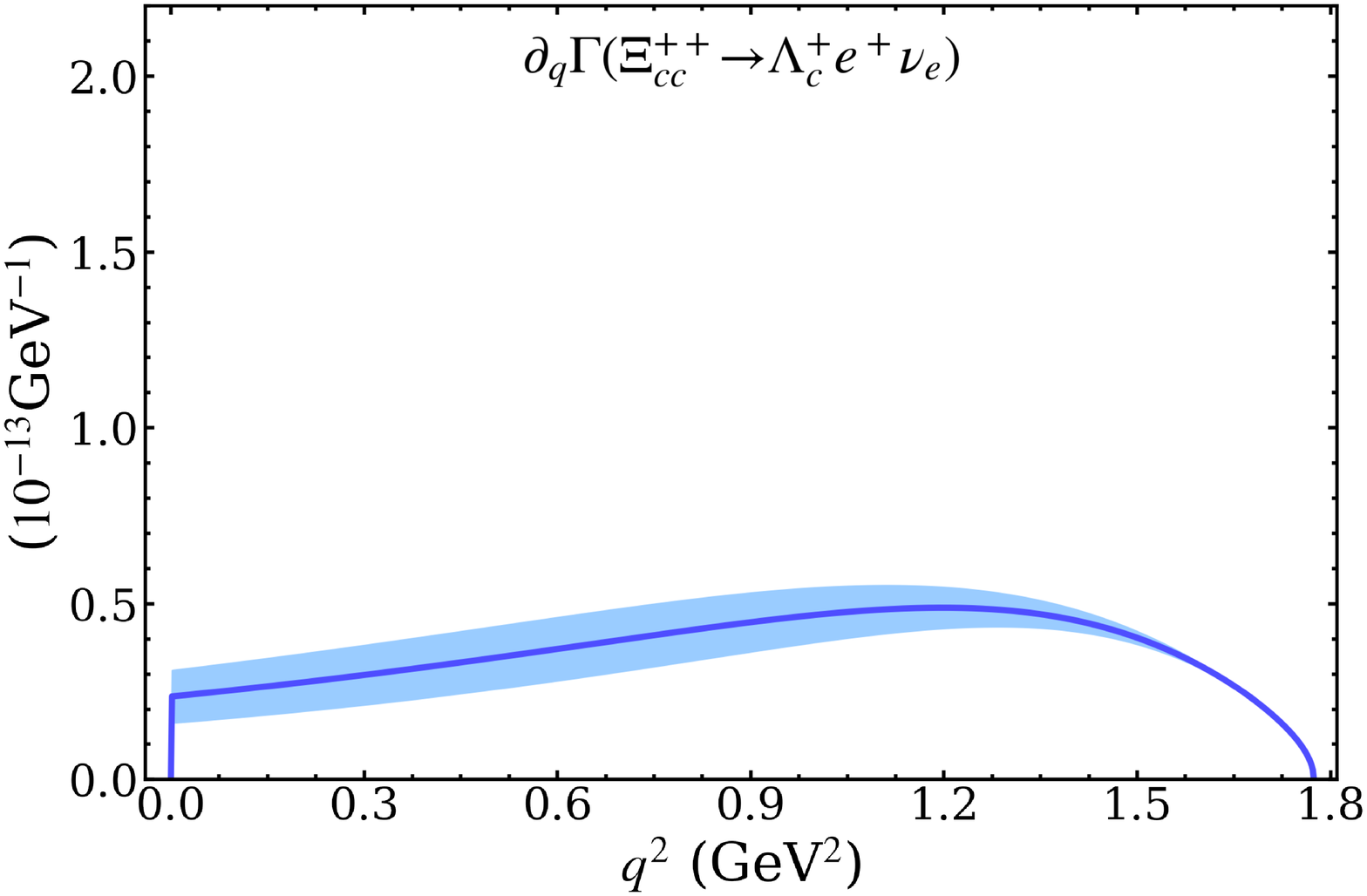}\label{fig: Omega decay widths-5}}\quad
    \subfloat[]{\includegraphics[width=.45\textwidth]{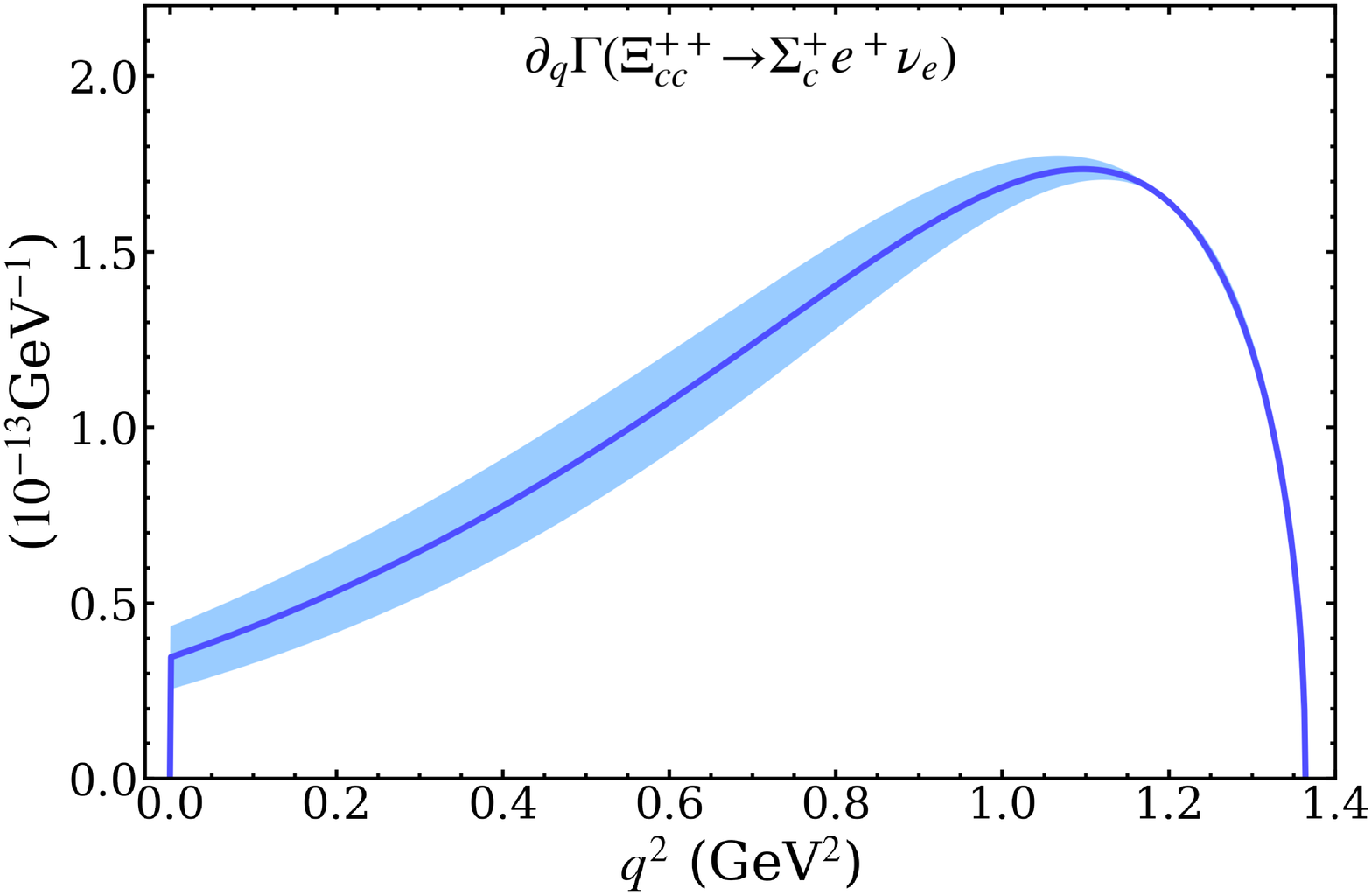}\label{fig: Omega decay widths-6}}
    
    \caption{\label{fig: Omega decay widths} The partial decay widths for $B_{cc}\rightarrow B_{c}e^{+}\nu_{e}$, where the dashed lines and band widths are the center values and uncertainties, while (\protect\subref*{fig: Omega decay widths-1}, \protect\subref*{fig: Omega decay widths-2}, \protect\subref*{fig: Omega decay widths-3}) and (\protect\subref*{fig: Omega decay widths-4}, \protect\subref*{fig: Omega decay widths-5}, \protect\subref*{fig: Omega decay widths-6}) correspond to the $c\to s$ and $c\to d$ transitions, respectively.}
\end{figure}

\begin{figure}[htb!]
  \centering
  \subfloat[]{\includegraphics[width=.45\textwidth]{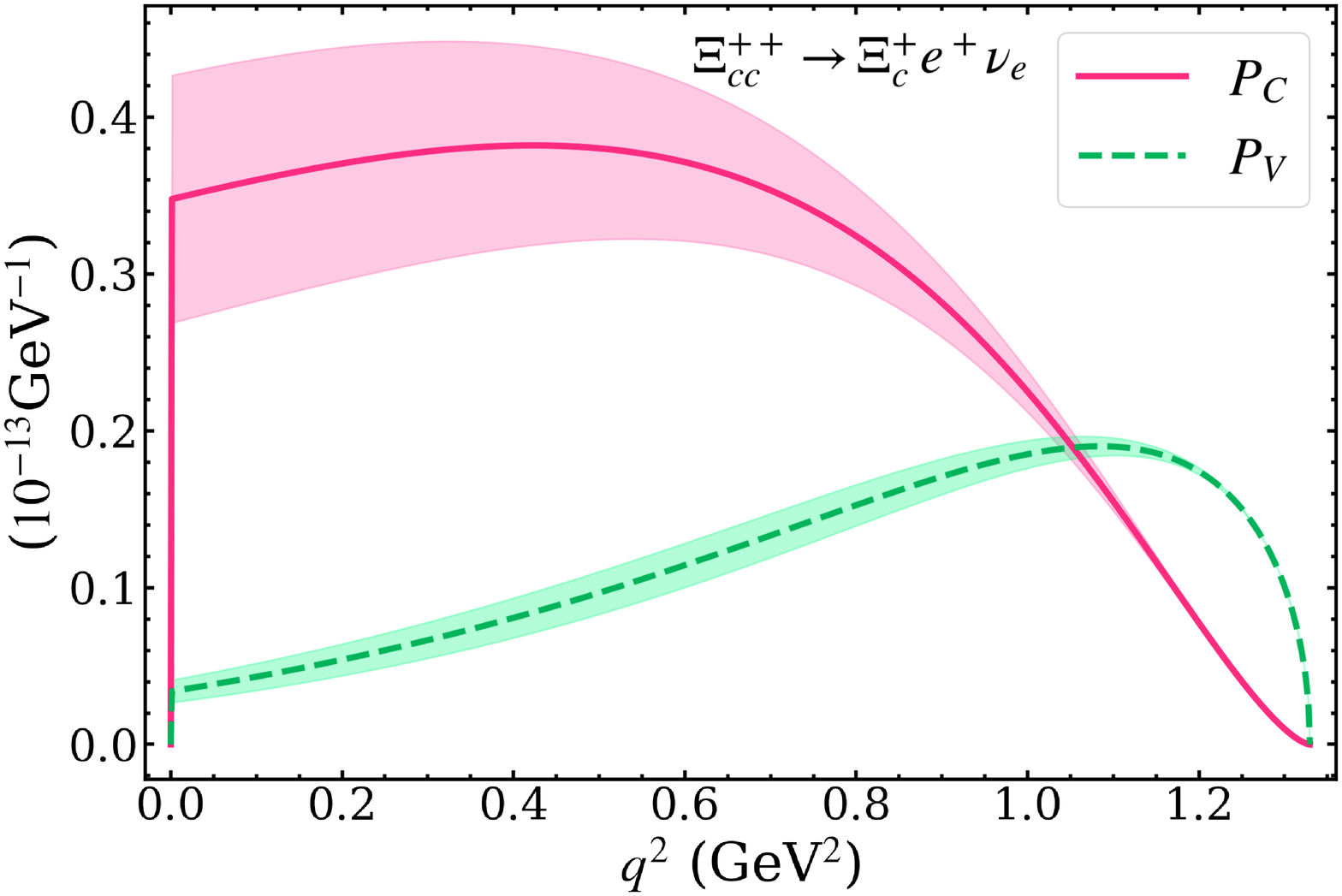}\label{fig: detail of differential decay widths-1}}\quad
  \subfloat[]{\includegraphics[width=.45\textwidth]{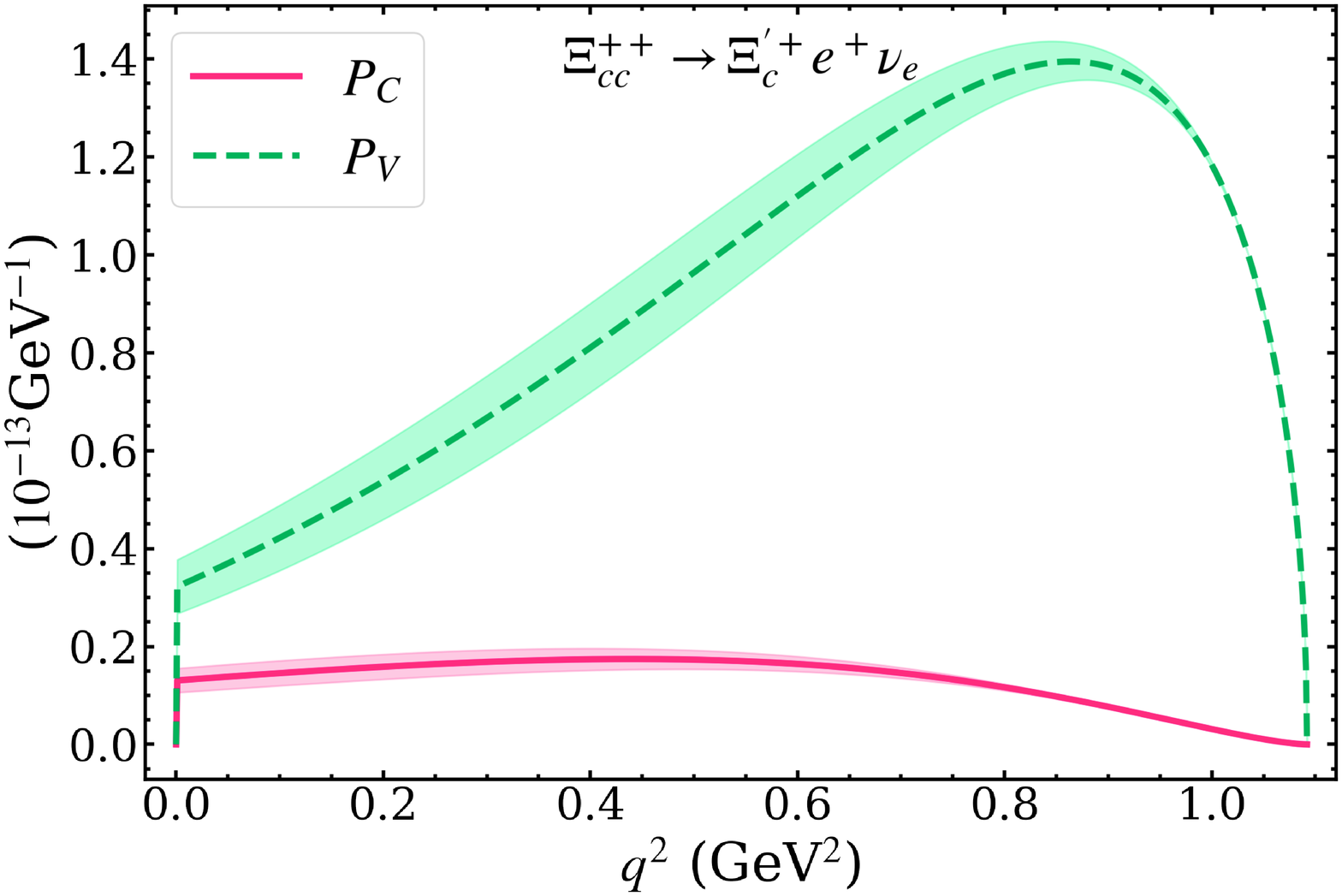}\label{fig: detail of differential decay widths-2}}
  
  \caption{\label{fig: detail of differential decay widths} The partial decay widths of $P_C$ and $P_C$ for $\Xi_{cc}^{++} \to \Xi_c^{(\prime) +} e^+  \nu_e$. 
  }
\end{figure}

In FIG.~\ref{fig: Omega decay widths}, we plot the differential decay widths. In the high $q^2$ areas, the uncertainties are minor since $f_1$ and $g_1$ have few errors, as explained at the beginning of this section.  
We see that $\partial_q\Gamma(\Xi_{cc}^{++} \to \Xi_c^+e^+\nu_e)$ is much smoother than the others. To see the underlying reason, 
we define the parity conserving and violating partial decay widths as 
\begin{equation}
    P_C\equiv         \partial_q \Gamma ( g_i=0)\,,\ P_V\equiv         \partial_q \Gamma ( f_i=0)\,,\         \partial_q \Gamma = P_C +P_V\,,
\end{equation}
where in $P_{C(V)}$ we take $g_{1,2,3}(f_{1,2,3})$ as zero, corresponding to the parity conserving (violating) parts of the decay widths.
We show $P_{C,V}(\Xi_{cc}^{++}\rightarrow\Xi_{c}^{(\prime)+})$ in FIG.~\ref{fig: detail of differential decay widths}.

We see that $P_{C,V}$ have very different behaviors. On the one hand, $P_C$ contribute to $\Gamma$ mostly in the low $q^2$ regions, and their values decrease quickly as $q^2$ goes up. On the other hand, $P_V$ behave oppositely. In $\Xi_{cc}^+ \to \Xi_c^+ e ^+  \nu_e$, the net result is that the tendencies of $P_{C,V}$ smear out each other in $ \partial_q \Gamma $. In contrast, the behavior of $\partial_q \Gamma$ is dominated by $P_V$ in $\Xi_{cc}^{++} \to \Xi_c^{\prime +} e ^+ v _e$. The sharp difference can be traced back to the spin-flavor overlappings, where we approximately have $P_C\propto N_{\text{unflip}}^2$ and $P_V\propto N_{\text{flip}}^2$. From TABLE~\ref{tab: Overlaps}, we see that $(N_{\text{flip}} / N_{\text{unflip}})^2 $ are $1/9$ and $100/36$ for $\Xi_{cc}^{++} \to \Xi_c^+$ and $\Xi_{cc}^{++} \to \Xi_c^{\prime +}$, respectively, which explains the opposite behaviors.

To test the lepton universality in the future experiments, we provide the ratios of 
$$\mathcal{R} = \Gamma(B_{cc}\rightarrow B_{c}\mu^{+}v_{\mu})/\Gamma(B_{cc}\rightarrow B_{c}e^{+}v_{e})$$ in Table~\ref{tab: decay widths for ratio}. 
Their values are close to but below 1.
Clearly, if $\mathcal{R}>1$ in the future experiment, it will be a signal of new physics.

\begin{table}
    \caption{\label{tab: decay widths for ratio} The ratios $\mathcal{R}$.}
    \centering
\begin{tabular}{cc|cc}
    \hline \hline
    $~c\to s~$& $~~~\mathcal{R}~~~$&$~c\to d~$&$~~~\mathcal{R}~~~$\\
    \hline
    $\Xi^{++}_{cc}\rightarrow \Xi^+_{c}$          & 0.99 & $\Xi^{++}_{cc}\rightarrow \Lambda^{+}_{c}$    & 1.00 \\
    $\Xi^{+}_{cc}\rightarrow \Xi^{0}_{c}$         & 0.99 & $\Omega^{+}_{cc}\rightarrow \Xi^{0}_{c}$        & 1.00 \\
    $\Xi^{++}_{cc}\rightarrow \Xi^{\prime +}_{c}$ & 0.97 & $\Xi^{++}_{cc}\rightarrow \Sigma^{+}_{c}$     & 0.97 \\
    $\Xi^{+}_{cc}\rightarrow \Xi^{\prime 0}_{c}$  & 0.97 & $\Omega^{+}_{cc}\rightarrow \Xi^{\prime 0}_{c}$ & 0.97 \\
    $\Omega^{+}_{cc}\rightarrow \Omega^{0}_{c}$   & 0.97 & $\Xi^{+}_{cc}\rightarrow \Sigma^{0}_{c}$     & 0.97 \\
    \hline \hline 
\end{tabular}
\end{table}

To eliminate the uncertainties caused by $V_{cf}$, by defining
\begin{equation}
  \text{G}_{B_{c}}^{B_{cc}}=\frac{1}{|V_{cf}|^{2}} \Gamma(B_{cc}\rightarrow B_{c}e^+  \nu_e),
\end{equation}
we find that
\begin{equation}\label{su3}
\begin{aligned}
\text{G}_{\Xi_{c}^{+}}^{\Xi_{cc}^{++}}:\text{G}_{\Xi_{c}^{0}}^{\Xi_{cc}^{+}}:\text{G}_{\Lambda_{c}^{+}}^{\Xi_{cc}^{++}}:\text{G}_{\Xi_{c}^{0}}^{\Omega_{cc}^{+}}&=1:0.99:1.24:1.17,\\
\text{G}_{\Xi_{c}^{\prime +}}^{\Xi_{cc}^{++}}:\text{G}_{\Xi_{c}^{\prime 0}}^{\Xi_{cc}^{+}}:\frac{1}{2}\text{G}_{\Omega_{c}^{0}}^{\Omega_{cc}^{+}}:\text{G}_{\Sigma_{c}^{+}}^{\Xi_{cc}^{++}}:\text{G}_{\Xi_{c}^{\prime 0}}^{\Omega_{cc}^{+}}:\frac{1}{2}\text{G}_{\Sigma_{c}^{0}}^{\Xi_{cc}^{+}}&=1:0.99:1.01:1.30:1.32:1.31,  
\end{aligned}    
\end{equation}
which are all expected to be 1 in the  exact  $SU(3)_F$ symmetry. 
As mentioned in the discussions of the form factors early, the main $SU(3)_F$ breaking effects come from the phase space difference. For instance, 
from FIG.~\ref{fig: Omega decay widths}, the phase space of $\Xi_{cc}^{++} \to \Lambda_c^+$ is 1.3 times larger than the one of $\Xi_{cc}^{++} \to \Xi_c^+$, which partly explains the ratios in Eq.~\eqref{su3} . 

\begin{table}
\caption{\label{tab: decay widths} The decay widths in units of $10^{-14}$ GeV along with the ones in the literature.}
\centering

\begin{threeparttable} 
\resizebox{\textwidth}{!}
{\begin{tabular}{lp{2.5cm}<{\centering}p{2.5cm}<{\centering}p{2.5cm}<{\centering}p{2.5cm}<{\centering}p{2.5cm}<{\centering}p{2.5cm}<{\centering}}
\hline
$B_{cc}\rightarrow B_{c}e^+ \nu_e$&This work &MBM~\cite{Perez-Marcial:1989sch} &HQSS~\cite{Albertus:2011xz} &LFQM~\cite{Wang:2017mqp} & LFQM~\cite{Hu:2020mxk} &QCDSR~\cite{Shi:2019hbf}  \\
\hline 
\hline
$\Xi^{++}_{cc}\rightarrow \Xi^+_{c}e^+ \nu_e$           & $5.11\pm 0.64$ & 7.36  & 5.78 & 11.50&8.74  &$7.72\pm 3.70$            \\
$\Xi^{+}_{cc}\rightarrow \Xi^{0}_{c}e^+ \nu_e$          & $5.08\pm 0.64$ & 7.36  & 5.73 & 11.40&8.63  &$7.72\pm 3.70$            \\
$\Xi^{++}_{cc}\rightarrow \Xi^{\prime +}_{c}e^+ \nu_e$  & $10.92\pm 0.81$& 17.56 & 9.64 & 12.80&14.30 &$5.31\pm 3.52$            \\
$\Xi^{+}_{cc}\rightarrow \Xi^{\prime 0}_{c}e^+ \nu_e$   & $10.85\pm 0.81$& 13.02 & 9.57 & 12.70&14.10 &$5.31\pm 3.52$            \\
$\Omega^{+}_{cc}\rightarrow \Omega^{0}_{c}e^+ \nu_e$    & $22.09\pm 1.63$& 26.76 & 18.61& 25.50&28.00 &$12.50\pm 8.02$           \\
\hline 
$\Xi^{++}_{cc}\rightarrow \Lambda^{+}_{c}e^+ \nu_e$     & $0.34\pm 0.06$ & 0.46  & 0.32 & 1.05 &0.80  &$0.76\pm 0.37$            \\
$\Omega^{+}_{cc}\rightarrow \Xi^{0}_{c}e^+ \nu_e$       & $0.32\pm 0.04$ & 0.46  & 0.27 & 0.81 &0.59  &$0.61\pm 0.28$            \\
$\Xi^{++}_{cc}\rightarrow \Sigma^{+}_{c}e^+ \nu_e$      & $0.76\pm 0.06$ & 0.78  & 0.52 & 0.96 &1.09  &$0.49\pm 0.29$            \\
$\Omega^{+}_{cc}\rightarrow \Xi^{\prime 0}_{c}e^+ \nu_e$& $0.77\pm 0.06$ & 0.91  & 0.49 & 0.93 &1.03  &$0.56\pm 0.35$            \\ 
$\Xi^{+}_{cc}\rightarrow \Sigma^{0}_{c}e^+ \nu_e$       & $1.52\pm 0.12$ & 1.50  & 1.04 & 1.91 &2.17  &$0.99\pm 0.58$            \\
\hline
\hline
\end{tabular}}
\end{threeparttable}
\end{table}

The computed decay widths, along with those in the literature, are shown in Table~\ref{tab: decay widths}, where Ref.~\cite{Perez-Marcial:1989sch} computes the form factors by the MBM, Ref.~\cite{Albertus:2011xz} analyzes the decays  with  the heavy quark  spin symmetry~(HQSS), Refs.~\cite{Wang:2017mqp,Hu:2020mxk} adopt the LFQM with different sets of the parameter input, and Ref.~\cite{Shi:2019hbf} calculates the decay widths by the QCDSR. 
The decay width of $\Xi_{cc}^{++} \to \Xi_c^{(\prime)+} e^+\nu_e$ is slightly larger than the one of  
$\Xi_{cc}^{+} \to \Xi_c^{(\prime)0} e^+\nu_e$ as $M_{\Xi_{cc}^{++}}$ is 2 MeV larger than $M_{\Xi_{cc}^+}$.   
For the $c\to s$ transition, our results of the decay widths are well consistent with the ones of the HQSS but systematically lower than those in the LFQM~\cite{Hu:2020mxk} and MBM~\cite{Perez-Marcial:1989sch} by a factor of 1.5. 
We note that in the MBM~\cite{Perez-Marcial:1989sch},  the $q^2$ dependencies of the form factors are put by hand and independent of the spectator quarks. For instance, the $q^2$ dependencies of  $\Lambda_c  \to \Lambda$ and $\Xi_{cc} ^{++} \to \Xi_c^+$ are taken to be the same in the MBM~\cite{Perez-Marcial:1989sch}. However, we emphasize that the spectator effects of the charm quark and others are very different, as shown explicitly in $D_q^v$ of Eq.~\eqref{2.25}. The exponential factor of $\exp ( -2i E_c v z )$  deviates largely to those of $\exp ( -2i E_{u,d,s} v z )$. In particular, we find that it suppresses the decay widths  by  more than $40\%$, which causes the deviations between the results of the MBM and ours.

\section{Conclusion}\label{Conclusion}
We have studied the semileptonic decays of the doubly charmed baryons. Explicitly, we have found that $\Gamma(\Xi_{cc}^{++}{\rightarrow}\Xi_c^+e^+\nu_e, \Xi_c^{\prime+}e^+\nu_e, \Lambda_c^+e^+\nu_e, \Omega_c^+ e^+\nu_e) 
=(5.1\pm 0.1 , 11\pm 1, 0.34\pm 0.06, 0.76\pm 0.06)\times10^{-14}$~GeV, 
$\Gamma(\Xi_{cc}^+\rightarrow \Xi_c^0e^+\nu_e, \Xi_c^{\prime0}e^+\nu_e , \Sigma_c^0e^+\nu_e) = (
5.1\pm 0.6, 11\pm 1, 1.5\pm 0.1) \times10^{-14}$~GeV, 
and $\Gamma(\Omega_{cc}^+\rightarrow \Omega_c^0 e^+\nu_e, \Xi_c^0e^+\nu_e , \Xi_c^{\prime0} e^+\nu_e) = 
(22\pm 2, 0.32 \pm 0.04, 0.77\pm 0.06)\times10^{-14}$~GeV.  We have discussed the $SU(3)_F$ breaking effects regarding the aspects of (\romannumeral1) the phase spaces differences, (\romannumeral2) the spectator quark effects, and (\romannumeral3) the overlappings of the transited quarks. We have shown that the other breaking effects are negligible compared to the phase space differences. In particular, the form factors well respect the $SU(3)_F$ relations using $\omega$ as the variables. 
In addition, we have obtained that $\Gamma(\Xi_{cc}^{++} \to \Lambda_c ^+ e^+ \nu _e) V_{cs}^2/\Gamma(\Xi_{cc}^{++} \to \Xi_c ^+ e^+ \nu_e )V_{cd}^2  = 1.24$, which is expected to be $1$ under the exact $SU(3)_F$ symmetry. 

The behaviors of the parity violating and conserving partial decay widths have been examined. Accordingly, we have demonstrated that the partial decay width of $\Xi_{cc}^{++} \to \Xi_c^+ e^+  \nu_e$ is smoother than others, which can be testified in future experiments. 
We have also shown that the spectator effects of the charm quark suppress the decay widths by $40\%$ and shall not be taken to be the same as the other quarks. 

\acknowledgments

This work is supported in part by the National Key Research and Development Program of China under Grant No. 2020YFC2201501 and  the National Natural Science Foundation of China (NSFC) under Grant No. 12147103.

\end{document}